\documentclass[]{article}
\usepackage[utf8]{inputenc}
\usepackage[english]{babel}
\usepackage{amsmath}
\usepackage{amsfonts}
\usepackage{amssymb}
\usepackage{graphicx}
\usepackage{subcaption}
\usepackage{braket}
\usepackage{geometry}
\usepackage{cite}
\usepackage{authblk}

\title{Entanglement on curved hypersurfaces:\\ 
 A field-discretizer approach}
\author{Tal Schwartzman and Benni Reznik}
\affil{School of Physics and Astronomy, Tel-Aviv University, Tel Aviv 69978, Israel}

\begin{document}

\maketitle

\begin{abstract}
	
We propose a covariant scheme for measuring entanglement on general hypersurfaces in relativistic quantum field theory. 
For that, we introduce an auxiliary relativistic field, 'the discretizer', that by locally interacting with the field along a hypersurface, fully swaps the field's and discretizer's states. 
It is shown, that the discretizer can be used to effectively 
cut-off the field's infinities, in a covariant fashion, and without having to introduce a spatial lattice. 
This, in turn, provides us an efficient way to evaluate entanglement between arbitrary regions on any hypersurface. 
As examples, we study the entanglement between complementary and separated regions in 1+1 dimensions, for flat hypersurfaces in Minkowski space, for curved hypersurfaces in Milne space, and for regions on hypersurfaces approaching null-surfaces. 
Our results show that the entanglement between regions on arbitrary hypersurfaces in 1+1 dimensions depends only on the space-time endpoints of the regions, and not on the shape of the interior. Our results corroborate and extend previous results for flat hypersurfaces.

\end{abstract}

\section{Introduction}

There has been recently much interest in exploring the nature of  "quantum information"  and entanglement
 in the framework of relativistic quantum field theories.    
From the perspective of quantum information,   the relativistic set up has several appealing features.
The relativistic framework comes with  a  built-in causal structure  that provides a concise meaning to the so called set of  "local operations",  
that keep entanglement invariant,   and used to manipulating and detect entanglement. 
Yet another,  perhaps more striking feature,    in which the relativistic case differs from to the non-relativistic, 
follows from the generalized evolution discovered by 
Tomonaga and Schwinger. 
While,  non-relativistically states are defined on,   and propagate between,   surfaces  of  common time-"t",
in relativistic quantum field theories,      quantum states can be assigned on a general Cauchy hypersurfaces.
Temporal evolution then propagates states along a family of hypersurfaces that foliate space time. 
In relativistic quantum field theory we can study  in principle, the entanglement structure  with respect to some division into regions on a general hypersurface. For a state specified on the slice $t=t_0$, one can nevertheless ask how entangled are different regions, 
if experiments are done on the space-like hypersurface 
$t=t(x)$.

The need to study entanglement on general hypersurfaces,   seems to appear naturally  in connection with quantum black holes. 
Black-holes carry entropy,  that at least in part,  is accounted by,  the so called  entanglement-entropy,  between degrees of freedom
residing at interior and exterior regions.   The relation between entanglement,  black hole entropy,   and the information problem, 
 has been studied  extensively.
\cite{BombelliEntropyBlackHoles:1986, SrednickiEntropyArea:1993, HOLZHEY1994443,PolchinskiParadox}. 

In general, it is difficult to compute entanglement in quantum field theory without imposing certain simplifications. Partly because, a field carries infinitely many degrees of freedom, one at each point in space, which results in divergences. 
To control the infinities, several methods have been developed and used over the years.
a) Methods that impose some sort of discretization, typically, by converting the system to a spatial lattice, while retaining time continuous~\cite{BombelliEntropyBlackHoles:1986, SrednickiEntropyArea:1993, SpatialStructBotero:2004, CriticalAndNonCriticalReznik:2009, AreaLawsPlenio:2010}.
b) Methods that build on the path integral~\cite{Calabrese_2004, Casini_2009, EntanglementNegativityinQuantumFieldTheory:2012, Calabrese_2013}. 
c) And methods that use, two or more, Unruh-DeWitt (UDW) detectors to extract vacuum entanglement~\cite{Reznik:2003aa, reznik2000distillation, ViolatingBellsInequalities:2005, Massar2006, Menicucci:2013, HarvestingFromVacuumMartinz:2015, Kempf:2016}.
The persistence of bipartite and multipartite vacuum entanglement, between arbitrary separated regions has been demonstrated in  
~\cite{ViolatingBellsInequalities:2005, SilmanReznikManyRegion:2005, SilmanReznikLongRangeDirac:2007, CriticalAndNonCriticalReznik:2009, ExtractionTripartite:2014,BerKennethReznik}.

For the present purpose of studying entanglement on curved hypersurfaces, each method has its own drawback: a) Spatial discretization methods impose a particular cut-off, which breaks the Lorentz invariance
and the causal structure of the theory~\cite{Peres:1983aa}. This means that although space like separated field operators commute at $t=t_0$, this is no longer the case on $t=t(x)$ as the evolution in the discrete theory is not causal. By imposing a particular discretization on $t=t_0$, one loses the notion of locality and the tensor product structure on $t=t(x)$.
b) Path integral methods can be difficult for numeric calculations, and analytical results are limited to particular cases.
c) Finally, while entanglement probes, can be used in curved space-times~\cite{Menicucci:2009, Fuentes:2010, KempfWeakGrav:2011, Nambu:2013, Nambu:2017,  Martin-Martinez:2014aa, SpaceTimeStructMartinz:2016, Mann:2018aa, MannMartinz:2018, Mann:2019aa, KempfMartinz:2020}, it is not clear how they can be used for a general hypersurface.
Furthermore, since they rely on point-like probes, that are unable to account
for the full field entanglement, but rather provide only lower bounds. 
In 1+1 dimensions, for instance, 
given by two regions of size $R$  and separation $L$,  
the detectors provide the bound $E_N\ge \textrm{e}^{-(L/R)^3}$~\cite{ViolatingBellsInequalities:2005}, 
while other methods find  $E_N \propto \textrm{e}^{-c {L/R}}$, where $c\sim 2\sqrt{2}$~\cite{CriticalAndNonCriticalReznik:2009, EntanglementNegativityinQuantumFieldTheory:2012}.

In the present work, we propose a new approach to entanglement on general hypersurfaces, that overcomes some of the eluded drawbacks of methods a) and c). 
For that, we introduce an auxiliary relativistic field, 'the discretizer', that generalize the probe method (c). 
We show that by locally interacting with the field along a hypersurface, we  can fully swap (instantaneously in the ideal case) the field's and discretizer's states. 
Furthermore, we show that the discretizer field can be used to effectively cutoff infinities, in a covariant fashion, without having to introduce a spatial cutoff. 
This, in turn, provides us with an efficient method for evaluating the entanglement between arbitrary regions on any hypersurface.


\section{Quantum states  on a hypersurface} \label{sec:QStates}

Consider a scalar field $\phi(z)$ 
in a $d+1$-dimensional space-time with a metric 
$g_{\mu\nu}(z)$
and  an action  
\begin{align}
S_\phi =\int d^{d+1}z \frac{1}{2}\sqrt{|g(z)|}\left(g^{\mu \nu}(z) \partial_\mu \phi(z) \partial_\nu \phi(z) -  V(\phi) \right).
\end{align}

\begin{figure}[h!]
	\centering
	\includegraphics[width=0.5\linewidth]{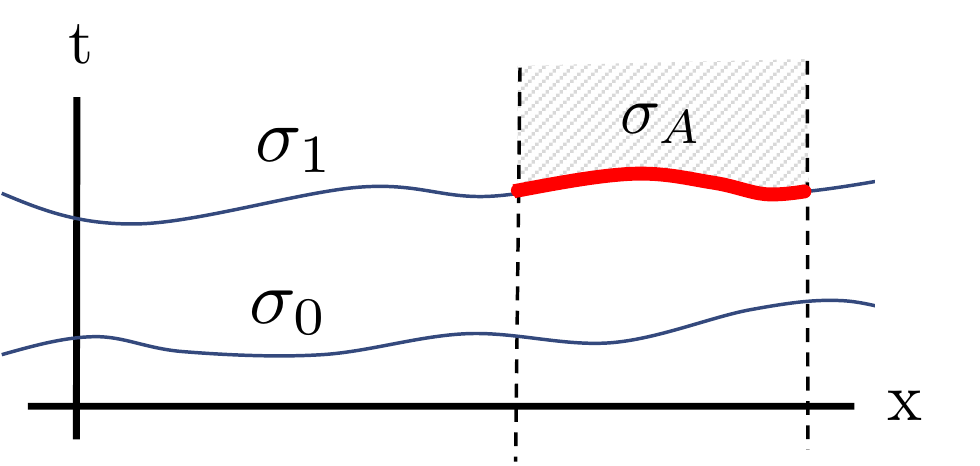}
	\caption[]{Consider the coordinates $t,x$, with some metric, that for example can be the Minkowskian $\eta_{\mu\nu}$. The state of the system can be defined on the hypersurfaces, $\sigma_0$ and $\sigma_1$, and we can find a "time" $\tau$ that evolves us between them, such that when $\tau(t,x) = \tau$, $\sigma_\tau$. The impulsive coupling acts in the hyper-region $\sigma_A$, which is part of $\sigma_1$}
	\label{fig:plotRegionOnCurved}
\end{figure}

The action is invariant under arbitrary coordinate transformations
$z \rightarrow z'(z)$,   that transform  the metric as a tensor according to $g_{\mu\nu} \rightarrow g_{\mu'\nu'}= z^{\mu}_{,\mu'}z^{\nu}_{,\nu'}g_{\mu\nu} $ ,  with $z^{\mu}_{,\mu'} =\partial z^{\mu}/\partial z^{\mu'}$.
This essentially means that we are free to choose a coordinates system such that our original metric takes the  form 
 $ds^2 = g_{\mu\nu}dz^\mu dz^\nu =g_{0 0} (z) d\tau^2 +2 g_{0 i}(z)d\tau dz + g_{ij} (z) dz^idz^j$
where $\tau=\tau_0$ defines hypersurfaces that foliate space-time.   It is possible therefore to examine how the system evolves in "time" through a one-parameter family of Cauchy surfaces $\sigma_\tau$ (See Fig. \ref{fig:plotRegionOnCurved} for an illustration). Each such surface $\sigma_\tau$ carries $d$ coordinates and a space-like metric $g_{ij}(z)$. 

We notice that freedom is non-trivial even when
space-time is globally flat and admits in some coordinates system a  Minkowski metric 
$ds^2= d\tau^2- d\boldsymbol{z}^2=\eta_{\mu\nu}dz^\mu dz^\nu$. 
We can still choose a coordinate system that transforms 
 $\eta_{\mu\nu} \rightarrow g_{\mu\nu}$. 
 We can then equally well consider the system's evolution along the resulting  foliation determined by the transformed  metric.
 
 To discuss quantum states,   lets next consider the 
 Hamiltonian formalism of  quantum field theory.   The general covariance of the theory means that one can define
 a quantum state of the system $\ket{\Omega,\sigma_\tau}$, 
along a general hypersurface $\sigma_\tau$.  
Such states  can then evolve in "time" 
  through a sequence of surfaces as has been 
 shown  by Tomonaga and Schwinger. 

To canonically quantize the field on a general surface 
$\sigma_\tau$ we 
choose at each point on the surface a
 conjugate momentum  as
  $\pi(z)=\delta{S_\phi}/\delta(\partial_\tau\phi(z))$.   This gives
\begin{align}
\pi(z) = \sqrt{|g(z)|} g^{0 \mu}(z) \partial_\mu \phi(z) = \sqrt{|g(z)|} g^{\nu \mu}(z)\partial_\nu\tau \partial_\mu \phi(z), \label{eq:pi2}
\end{align}
which is a directional derivative normal to $\sigma_1$.
The scalar field and conjugate momenta then satisfy  
the usual commutation relation
\begin{align}
& [\phi(\tau,\boldsymbol{z}),\phi(\tau,\boldsymbol{z'})] = 0, \nonumber \\
& [\phi(\tau,\boldsymbol{z}),\pi(\tau,\boldsymbol{z'})]=i\delta(\boldsymbol{z-z'}), \label{eq:comRelation} \\ 
& [\pi(\tau,\boldsymbol{z}),\pi(\tau,\boldsymbol{z'})] = 0. \nonumber
\end{align}

Now let us assume that the initial quantum state, $\ket{\Omega, \sigma_0}$, has been specified on a 
particular hypersurface $\sigma_0$.   For simplicity,  if possible,  it is convenient to choose this state on a flat hypersurface with a Minkowski metric,  wherein the time evolution is simple, but this is not necessary.   The state can essentially be described by the set of all  correlations functions,  although a simple relations between correlations and states exists only for Gaussian states. 

Suppose that we wish to make predictions about the state at a different time slice, $\sigma_1$. For this purpose, we find a different foliation of space time such that at $\tau=\tau_0$, the time-slice is $\sigma_0$, and at $\tau=\tau_1$ it is $\sigma_1$. With that we can write the action as $S_\phi = \int d\tau L$ and find the Hamiltonian $H_t=i\partial_\tau$ that will generate the evolution to $\ket{\Omega,\sigma_1}$.

It is helpful to consider the evolution in the Heisenberg picture. By solving the field's equations of motion $\nabla_\mu\nabla^\mu \phi = -V'(\phi)$,  where  $\nabla_\mu$ denotes the covariant derivative with respect to the metric, it is possible to express the field on $\sigma_1$ (and anywhere in spacetime) as a function of the field and its conjugate momentum on $\sigma_0$. This enables the evaluation of all correlation functions on $\sigma_1$.

The Heisenberg picture provides a simplification. In an appropriate coordinate system, the equations of motion are easily solved. 
Suppose that for the coordinates $t,\boldsymbol{x}$ we know the two point correlation $\bra{\Omega,\sigma_0}\phi(t,\boldsymbol{x})\phi(t',\boldsymbol{x}')\ket{\Omega,\sigma_0}$ everywhere in spacetime. We can express the two point correlation function on $\sigma_1$ by restricting this expression to $\tau(t,\boldsymbol{x})=\tau_1$.
For example, let the state $\ket{\Omega_M,\sigma_0}$ be the Minkowski vacuum and let the metric in terms of $x^\mu$ be flat. The correlations on $\sigma_1$ are obtained from $\bra{\Omega_M,\sigma_0}\phi\left(x^\mu(\tau_1,\boldsymbol{z})\right)\phi\left(x^\mu(\tau_1,\boldsymbol{z}')\right)\ket{\Omega_M,\sigma_0}$.

It is important to note that on this hypersurface, the conjugate momentum that will respect the commutation relation in Eq. \eqref{eq:comRelation}, is not $\partial_{t}\phi$ but as defined in Eq. \eqref{eq:pi2}.
We can now compute correlations
involving conjugate momenta and/or fields on a general  hypersurface.   
 For instance for two points on $\sigma_\tau$ we get
\begin{align}
\langle  \phi(z) \pi(z') \rangle = \sqrt{|g|} g^{\nu \mu}\partial_\nu \tau \partial_{\mu}\langle 
\phi(z)\phi(z')\rangle.
\end{align}

Clearly this provides us by means to evaluate  any correlation function involving higher moments of the conjugate momenta and field.   In the simplifying case that the initial state is Gaussian and that the evolution operator is a Gaussian unitary, the state on any future or past hypersurface is also Gaussian. 
In this special case, the entanglement is determined by two-point functions of the form $\braket{\phi(z) \phi(z')}, \braket{\phi(z) \pi(z')},\braket{\pi(z) \pi(z')}$. 
In general, higher order correlations will be needed as well. 

The conjugate field and momenta on the curved hypersurface inherit the usual tensor product structure for states in separated regions.  Therefore the same means for computing entanglement can be used for a state on a curved hypersurface.
Unfortunately,  one encounters as in the flat case
difficulties involving field infinities.

\section{Swapping fields on a hypersurface} \label{sec:swap}

In order to extract information about the field's state and to evaluate its entanglement we introduce an auxiliary probe field $\psi'$ that will interact with the field. Unlike the probe methods mentioned in the introduction, 
the auxiliary field $\psi'$ interacts with the field $\phi$ instantaneously on a subregion of a single hypersurface, in relativistic covariant fashion. 

Let us introduce a second scalar field $\psi'$ with the same action as of the field discussed in the previous section:
$S_\psi' =S_{\phi \leftrightarrow \psi}$ .  We can define for it as discussed, canonical operators along a general hypersurface $\sigma_\tau$ such that $[\psi'(\tau,\boldsymbol{z}),\pi'(\tau,\boldsymbol{z'})]=i\delta(\boldsymbol{z-z'})$.

Let us consider a unitary operator $U$
\begin{align} \label{swapOperator}
U(\alpha) = \mathrm{e}^{-i\alpha\int d^d z\left[\phi(\tau_1,\boldsymbol{z})\pi'(\tau_1,\boldsymbol{z})-\pi(\tau_1,\boldsymbol{z}) \psi'(\tau_1,\boldsymbol{z})\right]}.
\end{align}
It can easily be verified that $U(\alpha)$ is a rotation operator. With the choice $\alpha = \pi/2$, $U$ transforms the operators on $\sigma_1$ as
\begin{align}
& \psi'(\tau_1,\boldsymbol{z}) \rightarrow \phi(\tau_1,\boldsymbol{z}), \label{eq:swap1} \\
& \pi'(\tau_1,\boldsymbol{z}) \rightarrow \pi(\tau_1,\boldsymbol{z}), \label{eq:swap2}
\end{align}
In the Shr\"odinger picture,  let us denote the states prior to the interaction by $ \Psi[\phi]$  and 
$ \Psi' [\psi']$,  the initial states of $\phi$ and $\psi'$  respectively.  Then the swap acts on the states as  
\begin{align} 
& \Psi'[\psi'] \rightarrow \Psi[\psi'] \\ &  \Psi[\phi] \rightarrow \Psi' [-\phi]
\end{align}

The unitary transformation in (\ref{swapOperator})
can be generated by an impulsive  interaction
 that is non-zero only on a particular hypersurface. 
 The corresponding action takes the covariant form
\begin{align}
S_{int} =  \int_{\sigma} d^{d+1} z \sqrt{-g}(\frac{1}{2}\epsilon^\mu \epsilon_\mu(\phi^2+\psi'^2)-\epsilon^\mu(\phi\partial_\mu \psi'
-\psi'\partial_\mu\phi)) \label{eq:action}
\end{align}
where  $\epsilon^\mu = \frac{\pi}{2} \delta(\tau - \tau_0)n^\mu$ has support only in $\sigma_{1}$,  and $n^\mu = g^{\mu \nu}\partial_\nu \tau$ being a vector normal to the hypersurface.    The fields  
interacts along
$\sigma_1$,  but otherwise,  prior to and subsequent to the interaction, remain uncoupled.
One can easily check that the action gives rise in the Hamiltonian formalism to an interaction term 
\begin{align}
H_{int} = \frac{\pi}{2}\delta(\tau-\tau_0)\int d^n z (\phi\pi'-\pi \psi').
\end{align}
which generates the unitary swap transformation in Eq. \eqref{swapOperator}.

In passing,  it is worth mentioning that the spring-like terms  $\propto (\phi^2 + \psi'^2)$ in the interaction Lagrangian,  have been first discussed by Bohr and Rosenfeld,  in their classic work on the measurability of the electromagnetic field~\cite{Bohr:1933zz, AharonovsBook:2003}.   Their physical significance  can be understood as follows: consider for simplification the interaction on a flat hypersurface, with the flat metric time $t$.
when we couple to the conjugate momentum $\pi$, we necessarily disrupted the equality of $\pi$ to the velocity $\partial_t\phi$,  while the interaction is ``on'':  $\pi \neq \partial_t\phi$. It can be easily verified, that the spring term, acts to restore this relation,
so that the measurement detects the undisturbed value of $\partial_t \phi$, that the field would have had, in an undisturbed evolution. 
Adding the spring-like "compensation" terms to the system will amend it to be $\delta \partial_t\Phi(x) = \partial_t\phi(x) + g(x)\Phi(x)$, which is constant and equals to $\partial_t\phi(x)$ just before the measurement.

Returning to our main issue,   let us discuss the effect of the 
swap on entanglement,   when the interaction takes place
only in the sub-region 
$\sigma_A\in \sigma_1$,   as described in Fig. 1.
In this case, we choose $\epsilon^\mu$ from Eq. \eqref{eq:action} to have support only on $\sigma_A$, thus the interaction swaps the states of the two fields 
in this region alone.   After the swap,  the $\phi$-field previous entanglement between regions $A=\sigma_A$ and its complement $\bar{A} =\sigma_1-\sigma_A$,   
is encoded as an entanglement 
of the $\psi'$-field in $A$ and $\phi$ in $\bar{A}$. 

To illustrate this,   let us decompose the field states,  by means of a Schmidt decomposition with respect to the regions $A$ and $\bar{A}$, as $\sum c_\phi |\phi(\sigma_A)\rangle |\phi(\sigma_{\bar{A}})\rangle$.  Consider next an ancillary field that vanishes outside  $\sigma_A$  in an initial (arbitrary) $|\psi(\sigma_A)\rangle'$, 
and consider an interaction that swaps  $|\psi(\sigma_A)\rangle'|\phi(\sigma_A)\rangle \to |\phi(\sigma_A)\rangle'|\psi(\sigma_A)\rangle$ (The apostrophe marks that the state belongs to the Hilbert space of the discretizer). Then the combined state evolves according to  
\begin{align} 
|\psi(\sigma_A)\rangle' \sum c_\phi|\phi(\sigma_A)\rangle |\phi(\sigma_{\bar{A}})\rangle \to
\sum  |\phi(\sigma_{A})\rangle' c_\phi|\psi(\sigma_A)\rangle |\phi(\sigma_{\bar{A}})\rangle. \label{eq:swapState}
\end{align}
Hence the field's entanglement becomes encoded in the correlations between the ancillary field in $\sigma_A$ and the system field degree of freedom not in $\sigma_A$. We note that the operator which realizes the swap is local in $\sigma_A$ and as such does not change the entanglement between $\sigma_A$ and $\sigma_{\bar{A}}$.

\section{Discretizer-field } \label{sec:Disc}

So far it may seem that we have just transformed our original problem to a similar one. In this section, we shall add constraints to the auxiliary field $\psi'$, which introduce discreteness to the field's modes. In the following, we will refer to the auxiliary field $\psi'$ as the "discretizer". 
The role of the discretizer will be twofold:  (a) To impose an effective cut-off on a general hypersurface, without breaking causality.
And (b), to map the system's continuous spectrum into a discrete one, thus providing simpler means to compute entanglement.

The dashed lines in Fig. 1 denote  boundary  conditions for the discretizer:  $\psi'(\text{boundary})=0$.
After the correlations are transformed between the discretizer and the system field, the two systems remain decoupled.
Therefore, a subsequent measurement of the discretizer, will not be restricted to $\sigma_A$, and can instead be carried out within the spacetime future domain of this section (denoted by a shaded region in Fig. \ref{fig:plotRegionOnCurved}).

We note that the discretizer
vanishes outside $\sigma_A$, and satisfies boundary conditions at $\partial\sigma_A$, and therefore can  be described by a discrete Hilbert space, between the modes confined in $\sigma_A$ and the field. 
The original field entanglement can therefore be expressed after the swap, in terms of an enumerable set of correlations, 
that can be used to implement a cut in mode space.
What previously was the system's state, is now the discretizer's state $|\phi(\sigma_{A})\rangle'$, which belongs in momentum space to a discrete tensor product Hilbert space,
\begin{align}
|\phi(\sigma_{A})\rangle' \in {\cal H}' =\bigg\lbrace {\cal H}_1\otimes\cdots \otimes {\cal H}_{N_c}\bigg\rbrace
\otimes{\cal H}_{N_c+1} \otimes{\cal H}_{N_c+2}\otimes \cdots
\end{align}
After the swap, the original field's entanglement between
$\sigma_A$ and its complement $\sigma- \sigma_A$, is now "stored" between the 
discretizer's degrees of freedom and, the field's degrees of freedom at $\sigma - \sigma_A$.
Hence the original fields entanglement can be 
expressed as a function of correlations between operators that belong to the discrete Hilbert space.
\begin{align}
E_{sys}(\sigma_A) \to E_D(\sigma_A) = \lim_{N_c\to \infty}  E_{N_c}({\cal H}_1\otimes\cdots \otimes {\cal H}_{N_c}), \label{eq:10}
\end{align}
where $ E_{N_c}$ stands for the contribution to entanglement, of all modes  up to $N_c$.

We notice that the set-up provides a simple way to impose a cut-off.
By truncating the limit in \eqref{eq:10}, we consider the entanglement contribution of only a finite set of modes, say $1\leq n\leq N_c$, in a sub-space denoted above by curly braces.
The highest mode, gives rise to a cut-off scale
which is controlled by $N_c$, therefore, rather then imposing the cut-off
in space (by discretizing the Hamiltonian), the cut-off will be imposed by tracing higher modes.

\section{Construction of the discrete modes} \label{sec:ConstuctModes}

Given the finite support of the discretizer, we can expand the discretizer field, prior to the swap, in terms of any set of normalized orthogonal functions ${h_n(\boldsymbol{z})}$, that satisfy the boundary conditions, and form a basis:
\begin{align}
& \psi'(\tau_0,\boldsymbol{z}) = \sum_n h_n(\boldsymbol{z}) \psi_n \ \ \ ; \ \ \ \psi_n = \int_{\sigma_A} d^n z h_n(\boldsymbol{z}) \psi'(\tau_0,\boldsymbol{z}), \label{eq:psiMode}
\end{align} 

To construct the conjugate pairs of each $\psi_n$, we find the conjugate momentum $\pi'(\tau,z)$ of $\psi'(\tau,z)$, for which $[\phi(\tau_0,z),\pi'(\tau_0,z')] = i\delta(z-z')$. With this in mind,
\begin{align}
\pi' (\tau_0,\boldsymbol{z}) = \sum_n h_n(\boldsymbol{z})\pi_{\psi_n}  \ \ \ ; \ \ \ \pi_{\psi_n} = \int_{\sigma_A} d^n z h_n(\boldsymbol{z}) \pi'(\tau_0,\boldsymbol{z}). \label{eq:piMode}
\end{align}
The discrete modes form a canonical set such that $[\psi_n,\pi_{\psi_m}] = i\delta_{n,m}$.
The local conjugate momentum $\pi' (\tau_0,\boldsymbol{z})$ is derived from the Lagrangian density as in Eq. \eqref{eq:pi2}.
For simplicity, in the following sections we choose a coordinate gauge $g_{0 i}=0$, such that the momentum has the form
$\pi(z)= \sqrt{|g(z)|} g^{\tau\tau}(z) \partial_\tau \phi(z)$.

\section{Evaluating the entanglement} \label{EvaluatingSec}

For evaluating the entanglement we now consider the discretizer after the swap took effect on $\sigma_A$.
As described in Eqs. \eqref{eq:swap1} and \eqref{eq:swap2},
the discrete modes transform as, 
\begin{align}
& \psi_n \to \int_{\sigma_A} d^n z h_n(\boldsymbol{z}) \phi(\tau_0,\boldsymbol{z}) \label{eq:psi_nTo} \\ &
\pi_{\psi_n} \to \int_{\sigma_A} d^n z h_n(\boldsymbol{z}) \pi(\tau_0,\boldsymbol{z}) = \int_{\sigma_A} d^n z h_n(\boldsymbol{z})\sqrt{|g(z)|} g^{\tau\tau}(z) \partial_\tau \phi(\tau_0,\boldsymbol{z}).\label{eq:pi_nTo}
\end{align}
The entanglement of the original field, 
can now be reached from the discretizer's discrete set of correlations, after the interaction.
By swapping the fields we manage to map the original continuous 
mode structure into a discrete one.
The computational essence of this method is that we express the field in the finite region in terms of a discrete set of local modes. The effect of boundary conditions and the use of local modes has been considered also in \cite{Vazquez:2014aa, Brown:2014qna}.

For the particular case of Gaussian states,  the evaluation turns out simple, since the entanglement can be determined from the covariance matrix $M$. We will mention here some key notions from the literature (e.g. \cite{Adesso_2007}) about continuous variable systems.  For the reduced state of $N$ modes of the discretizer,
\begin{align}
M_{i,j} = \braket{O_iO_j +O_jO_i} -2 \braket{O_i}\braket{O_j}, \label{eq:covMat}
\end{align}
where $O = (\psi_1,...,\psi_N,\pi_{\psi_1},...,\pi_{\psi_N})$. Assuming $\braket{O_i} = 0$, after the swap and for $i,j \leq N$:
\begin{align}
& M_{i,j} =M_{\psi_i \psi_j} = 2\int_{\sigma_A}\int_{\sigma_A}  d^n z_1 d^n z_2 h_i(\boldsymbol{z_1}) h_j(\boldsymbol{z_2}) \braket{\phi(\tau_0,\boldsymbol{z_1})\phi(\tau_0,\boldsymbol{z_2})}, \\ &
M_{N+i,N+j} = M_{\pi_{\psi_i} \pi_{\psi_j}} = 2\int_{\sigma_A}\int_{\sigma_A}  d^n z_1 d^n z_2 h_i(\boldsymbol{z_1}) h_j(\boldsymbol{z_2})  \braket{\pi(\tau_0,\boldsymbol{z_1}) \pi(\tau_0,\boldsymbol{z_2})}, \\ &
M_{N+i ,  j} = M_{\pi_{\psi_i} \psi_j} = \int_{\sigma_A}\int_{\sigma_A}  d^n z_1 d^n z_2 h_i(\boldsymbol{z_1}) h_j(\boldsymbol{z_2})  \braket{\{\pi(\tau_0,\boldsymbol{z_1}), \phi(\tau_0,\boldsymbol{z_2})\} },
\end{align}
where  $\pi(\tau_0,\boldsymbol{z_1})= \sqrt{|g(\tau_0,\boldsymbol{z_1})|} g^{\tau\tau}(\tau_0,\boldsymbol{z_1}) \partial_\tau \phi(\tau_0,\boldsymbol{z_1})$.

To calculate the entanglement between the $N$ modes to their environment, we need to find the symplectic eigenvalues $\nu_i$ of the covariance matrix . This can be done by a symplectic diagonalization (see e.g. \cite{Brown:2014qna})
or by finding the eigenvalues of the matrix $|i\Omega{M}|$, where $i\Omega_{ij} = [O_i,O_j]$ is the symplectic form. The diagonal covariance matrix will have the form of $\text{diag}(\nu_1,\nu_1,...,\nu_N,\nu_N)$. From these values we can compute the entanglement entropy:
\begin{align}
S_A = \sum_{k}\frac{\nu_k+1}{2}\log\left(\frac{\nu_k+1}{2}\right) - \frac{\nu_k-1}{2}\log\left(\frac{\nu_k-1}{2}\right). \label{eq:EntEntropy}
\end{align}

We also consider the entanglement between two regions $\sigma_A,\sigma_B\in\sigma_1$. Now there are two discretizers $\psi'^A$ and $\psi'^B$, one at each region. For each discretizer, its own set of orthonormal local mode functions $h^A_n(z)$ and $h^B_n(z)$, and the modes of discretizer $B$, $\psi^B_i,\pi^B_i$ are defined as in Eqs. \eqref{eq:psiMode} and \eqref{eq:piMode}, but with the integration done over region $\sigma_B$. The covariance matrix of the reduced state of $N$ modes of discretizer $A$ and $B$ will contain also correlations between the two regions. It is defined as in Eq. \eqref{eq:covMat}, but with $O = (\psi^A_1,...,\psi^A_N,\pi^A_{\psi_1},...,\pi^A_{\psi_N},\psi^B_1,...,\psi^B_N,\pi^B_{\psi_1},...,\pi^B_{\psi_N})$. 

For the entanglement between these regions, we use the logarithmic negativity \cite{RFWerner}, which can be evaluated with the symplectic eigenvalues of the partially transposed covariance matrix $\tilde{M}$. The partial transposition corresponds to a sign flip in the momentum of the transposed part, i.e. in Eq. \eqref{eq:covMat}, $O \to \tilde{O} = (\psi^A_1,...,\psi^A_N,\pi^A_{\psi_1},...,\pi^A_{\psi_N},\psi^B_1,...,\psi^B_N,-\pi^B_{\psi_1},...,-\pi^B_{\psi_N})$. From $\tilde{M}$'s symplectic eigenvalues $\tilde{\nu}$ we can calculate the logarithmic negativity: 
\begin{align}
E_\mathcal{N} = \begin{cases}
-\sum_{k}\log\tilde{\nu}_k,& \text{for } k:\tilde{\nu}_k<1.\\
0              & \text{if} \ \tilde{\nu}_i\geq 1 \ \forall \ i.
\end{cases}
\end{align}

To summarize the algorithm for Gaussian states: start by computing the system's field and momentum correlation functions, along $\sigma_1$. Then, choose a complete set of normalized orthogonal functions that satisfy the boundary conditions and form a basis. Compute the doubly smeared correlation functions which are the elements of the covariance matrix.  Find the symplectic eigenvalues for the entanglement evaluation. 

A final remark on the scheme: from the numerical calculations used in the present work, it appears that in order for the entanglement to be finite for $N$ modes, the conditions should be such that $h_n(z)$ vanish on the boundaries. Otherwise, another momentum cutoff for the system's field will possibly be needed. We suspect that this is because the entanglement divergence comes from the boundaries, therefore such conditions regularize it to be visible only when $N \to \infty$.

\section{Williamson modes} \label{modesShape}
After the symplectic diagonalization of the matrices $M$ and $\tilde{M}$ we are left with a new set of modes that have only self correlations, i.e. no cross terms in the covariance matrices. For the case of bi-partite pure state entanglement those modes are pairwise entangled to counterpart modes in the environment, with the entanglement entropy, given by Eq. \eqref{eq:EntEntropy} using only the single symplectic eigenvalue of the pair~\cite{ModewiseEntanglement:2003}. 
For the case of bi-partite mixed state entanglement such as separated regions in the vacuum, these new modes, as we will see in the following,  can tell us how to define new modes, one at region $A$ and one in $B$ such that measurements on those single modes will give the entanglement according to the single symplectic eigenvalue. 

The aforementioned implies that, a) From an experimental perspective, one can limit the measurements to a single mode which contributes most of the entanglement between the regions. This can simplify experiments from probing a large number of degrees of freedom to a single one. b) We can identify the spatial distribution of degrees of freedom in the regions, that contributes most to the entanglement. 

Let us elaborate.
The given eigenvectors of $|i\Omega{M}|$ and $|i\Omega\tilde{M}|$, can be treated as Fourier coefficients that multiply the $h_n(z)$ modes of the detector: If $u_n$ is an eigenvector, we can define $f_n(\boldsymbol z) = \sum_i (u_n)_i h_i( \boldsymbol z)$. By Fourier transforming the problem of diagonalizing $|i\Omega{M}|$ and $|i\Omega\tilde{M}|$, we get equations for finding the eigenfunctions $f_n(x), g_n(x)$ and their corresponding symplectic eigenvalues. From $|i\Omega{M}|$ we have:
\begin{align}
& \int_{\sigma_A} d \boldsymbol s\braket{\phi(\boldsymbol z) \phi(\boldsymbol s)} f_n(\boldsymbol s) = i{\nu_n\over 2}  g_n(\boldsymbol z) 
& \int_{\sigma_A} d\boldsymbol s\braket{\pi(\boldsymbol z) \pi(\boldsymbol s)} g_n(\boldsymbol s) = -i{\nu_n\over 2} f_n(\boldsymbol z), 
\end{align}
and from $|i\Omega\tilde{M}|$:
\begin{align}
& \int_{\sigma_A} d\boldsymbol s\braket{\phi(\boldsymbol z) \phi(\boldsymbol s)} f_n(\boldsymbol s) +\int_{\sigma_B} d\boldsymbol s\braket{\phi(\boldsymbol z) \phi(\boldsymbol s)} f_n(\boldsymbol s) = i{\tilde{\nu}_n\over 2} g_n(\boldsymbol z) \label{eq:27}  \\ 
& -\int_{\sigma_A} d\boldsymbol s\braket{\pi(\boldsymbol z) \pi(\boldsymbol s)} g_n(\boldsymbol s) + \int_{\sigma_B} d\boldsymbol s\braket{\pi(\boldsymbol z) \pi(\boldsymbol s)} g_n(\boldsymbol s) = 
\begin{cases}
i{\tilde{\nu}_n\over 2} f_n(\boldsymbol z) & \boldsymbol z \in \sigma_A \\ \\
-i{\tilde{\nu}_n\over 2} f_n(\boldsymbol z) &\boldsymbol  z \in \sigma_B,
\end{cases} \label{eq:28}
\end{align}
where $\phi(\boldsymbol z) \equiv \phi(\tau_0,\boldsymbol z)$.
The discrete version of these equations for an harmonic chain can be found in \cite{SpatialStructBotero:2004}.
It is interesting to note that these equations can be derived if we heuristically take the covariance matrix of the field to be in position space and treat it as a functional.

We can normalize the functions such that,  $\int dx f_n(x) g_n(x) =1$. These functions define the symplectic transformation that diagonalize the covariance matrix. In other words $\int dxf_n(x)\phi(x)$ and $\int dxg_n(x)\pi(x)$, where the integration is over both regions, are the Williamson modes. If we define
\begin{align}
&	q^A_n = \frac{\int_{\sigma_A} d\boldsymbol z f_n(\boldsymbol z) \phi(\boldsymbol z)}{\sqrt{\int_{\sigma_A}  d\boldsymbol z f_n(\boldsymbol z)g_n(\boldsymbol z)}}, \\ 
& 	p^A_n = \frac{\int_{\sigma_A}  d\boldsymbol z g_n(\boldsymbol z) \pi(\boldsymbol z)}{\sqrt{\int_{\sigma_A}  d\boldsymbol z f_n(\boldsymbol z)g_n(\boldsymbol z)}},
\end{align} 
we can define a new system of modes which $q_n^A, p_n^A,q_n^B,p_n^B$ are a part of, and the rest of the system's modes are defined such that all together, they respect the canonical commutation relations. Then, the reduced covariance matrix of only mode $n$ in region $A$ and $B$ will have the same symplectic eigenvalue $\nu_n$. This can be seen by expressing Eqs. \eqref{eq:27} and \eqref{eq:28} in terms of the new modes and comparing to the equations that are obtained for the symplectic eigenvalue of the reduced covariance matrix of mode $n$.

Therefore we can interpret $f_n(x)$ and $g_n(x)$ as the weight functions that specify the spatial structure of the modes that carry such entanglement between them. 
Given $f_n(x)$ and $g_n(x)$, we can perform measurements on just a single mode. In the appendix, a formulation for a relativistic von-Neumann measurement model. Adjusting Eq. \eqref{eq:appendMeas}, such that $f(x) = f_n(x)$ will give a measurement of this mode, can be done for the momentum mode with $g_n(x)$.

We comment that the modes $q^A_n$ and $p^A_n$, are in fact non-local objects in $\sigma_A$, that cannot be observed instantaneously (without violating causality). 
However, we can perform the measurements on the discretizer field after the swap. The discretizer field remains decoupled in the causal future of $\sigma_A$, and therefore the duration of the final measurement is not restricted in time. It is reasonable to suspect that we can realize such interaction by considering a local non instantaneous coupling with a single degree of freedom such as an oscillator.
We leave the problem of finding a realization for future work.

\section{Flat hypersurfaces} \label{sec:flat}
In this section we apply the discretizer method
in a $1+1$-dimensional space, and 
consider flat hypersurfaces, and an initial state 
of a Minkowskian vacuum. The metric is $ds^2 = dt^2-dx^2$, and the hypersurfaces are for constant $t$.  We shall compare our method to previously known results for the entanglement entropy of a single region and the logarithmic negativity between separated regions.
\begin{figure}[h!]
	\centering
	\includegraphics[width=0.5\linewidth]{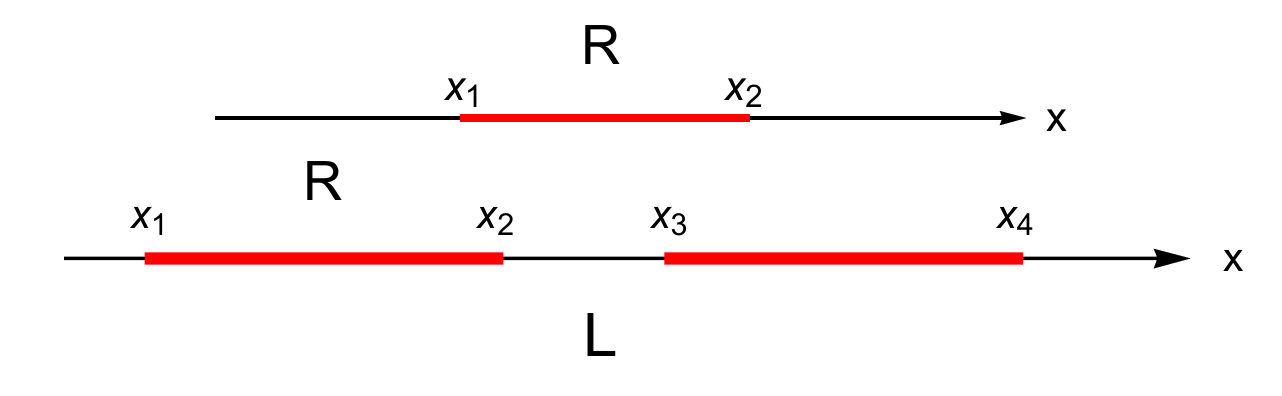}
	\caption[]{The discretizers are set in regions of size $R$ marked in red. The upper configuration is for measuring the entanglement between a single region and its environment and the lower configuration is for measuring the entanglement between two regions with distance $L$ between them.}
	\label{fig:plotRegionsFlat}
\end{figure}
We begin by expanding the discretizer 
in region $A=(x_1,x_2)$:
\begin{align}
& \ \psi'^A(x) = \sum_{n=1}^{\infty} \sqrt{{2}\over{R\omega_n}}\psi'^A_n \sin(k_n(x-x_1)) \ \ \ ; \ \ \  \psi'^A_n = \sqrt{{2\omega_n}\over{R}} \int_{x_1}^{x_2}dx\psi'^A(x)\sin(k_n(x-x_1)); \\ 
& \ \pi'^A(x) = \sum_{n=1}^{\infty} \sqrt{{2\omega_n}\over{R}}\pi'^A_n \sin(k_n(x-x_1)) \ \ \ \ ; \ \ \ \  \pi'^A_n = \sqrt{{2}\over{R\omega_n}} \int_{x_1}^{x_2} dx \pi'^A(x)\sin(k_n(x-x_1));
\end{align} 
where $k_n = {\pi n\over R}$, $R=x_2-x_1$ and $\omega_n = \sqrt{k_n^2 +M_0^2}$. Similarly, we expand the discretizer in region $B=(x_3,x_4)$, by changing the interval of integration and replacing  $\sin(k_n(x-x_1))$ in the equations above with $ \sin(k_n(x-x_3))$.

The evolution operator for the two regions 
is given in the interaction picture as \begin{align}
U = \mathrm{e}^{-i\frac{\pi}{2}\int_{A}dx(\phi\pi'^A-\pi\psi'^A)}\otimes\mathrm{e}^{-i\frac{\pi}{2}\int_{B}dx(\phi\pi'^B-\pi\psi'^B)}.
\end{align}
The effect of this operator on our field modes is given by:
\begin{align}
U^\dagger \psi'^A_n U = \psi^A_n = \sqrt{{2\omega_n}\over{R}} \int_{A}dx\phi(x)\sin(k_n(x-x_1)), \\
U^\dagger \pi'^A_n U  = \pi^A_n = \sqrt{{2}\over{R\omega_n}} \int_{A}dx\pi(x)\sin(k_n(x-x_1)), 
\end{align}
and similar transformation happen for the operators in region $B$. Suppose that both regions are of size $R$ and the separation between them is $x_3-x_2 = L$.

We can compute the CM's elements:
\begin{align}
& M_{\psi^A_n,\psi^B_m} =  {4\over R}\sqrt{\omega_n\omega_m}\int_{0}^{R}dx\int_{0}^{R}dy\sin({\pi n\over R}x)\sin({\pi m\over R}y)\int_{-\infty}^{\infty}{dk \over {4\pi\omega_k}}\mathrm{e}^{ik(x-y-L-R)}, \label{eq:34} \\
& M_{\pi^A_n\pi^B_m} = {4\over R\sqrt{\omega_n\omega_m}}\int_{0}^{R}dx\int_{0}^{R}dy\sin({\pi n\over R}x)\sin({\pi m\over R}y)\int_{-\infty}^{\infty}{dk\omega_k \over {4\pi}}\mathrm{e}^{ik(x-y-L-R)},\label{eq:35}
\end{align}   
where the case of a single region, $M_{\tilde{O}^A_n,\tilde{O}^A_m}$, is obtained by setting  $L+R = 0$ in Eqs. \eqref{eq:34} and \eqref{eq:35}. A change of variables $x \to Rx$ and $k \to k/R$ will give:
\begin{align}
& M_{\psi^A_n,\psi^B_m} =  \sqrt{\omega'_n\omega'_m}\int_{0}^{1}dx\int_{0}^{1}dy\sin({\pi n}x)\sin({\pi m }y)\int_{-\infty}^{\infty}{dk \over {\pi \sqrt{k^2+(M_0 R)^2}}}\mathrm{e}^{ik(x-y-{L\over R}-1)}, \\
& M_{\pi^A_n\pi^B_m} = \frac{1}{\sqrt{\omega'_n\omega'_m}}\int_{0}^{1}dx\int_{0}^{1}dy\sin({\pi n}x)\sin({\pi m}y)\int_{-\infty}^{\infty}{dk\sqrt{k^2+(M_0 R)^2} \over {\pi}}\mathrm{e}^{ik(x-y-{L\over R}-1)},
\end{align} 
where $\omega'_n = \sqrt{(\pi n)^2 + (M_0 R)^2}$. We can see that the integrals of the self correlations for a massless field are independent of $R$ so without introducing a physical scale like an energy cutoff the entanglement entropy is also independent of R. We can set $R = 1$ so that $L$ and $M_0$ are now dimensionless.
Doing the spatial integration we find
\begin{align}
& M_{\psi^A_n,\psi^B_m} = \sqrt{\omega_n\omega_m}2\pi n m\begin{cases}
\int {dk\over\omega_k} {2\cos^2({k\over2})\cos(k(1+L))\over (k^2-n^2\pi^2)(k^2-m^2\pi^2)} & n \  \text{ odd,} \ m \  \text{odd,} \\ \\
\int {dk\over\omega_k} {\sin({k})\sin(k(1+L))\over (k^2-n^2\pi^2)(k^2-m^2\pi^2)} & n \  \text{ odd,} \ m \  \text{even,} \\ \\
\int {dk\over\omega_k} {-\sin({k})\sin(k(1+L))\over (k^2-n^2\pi^2)(k^2-m^2\pi^2)} & n \  \text{ even,} \ m \  \text{odd,} \\ \\
\int {dk\over\omega_k} {2\sin^2({k\over2})\cos(k(1+L))\over (k^2-n^2\pi^2)(k^2-m^2\pi^2)} & n \  \text{ even,} \ m \  \text{even,} 
\end{cases} 
\end{align} 
The integrals can be further simplified to be more numerically tractable, 
by contour integrating in the complex plane. This gives:
\begin{align}
\nonumber M_{\psi^A_n,\psi^B_m} = & -\sqrt{\omega_n\omega_m}2\pi n m
\int_{M}^{\infty} {dy\over\sqrt{y^2-M^2}} {(-1)^n\mathrm{e}^{-y|L|} +(-1)^m\mathrm{e}^{-y(L+2)} -(1+(-1)^{n+m})\mathrm{e}^{-y(L+1)}\over (y^2+n^2\pi^2)(y^2+m^2\pi^2)} \\  & + \delta_{n,m}
\end{align}
\begin{align}
\nonumber M_{\pi^A_n,\pi^B_m} = & -{2\pi n m\over \sqrt{\omega_n\omega_m}}
\int_{M}^{\infty} {dy\sqrt{y^2-M^2}} {(-1)^n\mathrm{e}^{-y|L|} +(-1)^m\mathrm{e}^{-y(L+2)} -(1+(-1)^{n+m})\mathrm{e}^{-y(L+1)}\over (y^2+n^2\pi^2)(y^2+m^2\pi^2)} \\ & + \delta_{n,m}
\end{align}

\begin{figure}[h!] 
	\centering
	\begin{subfigure}[t]{0.48\linewidth}
		\includegraphics[width=\linewidth]{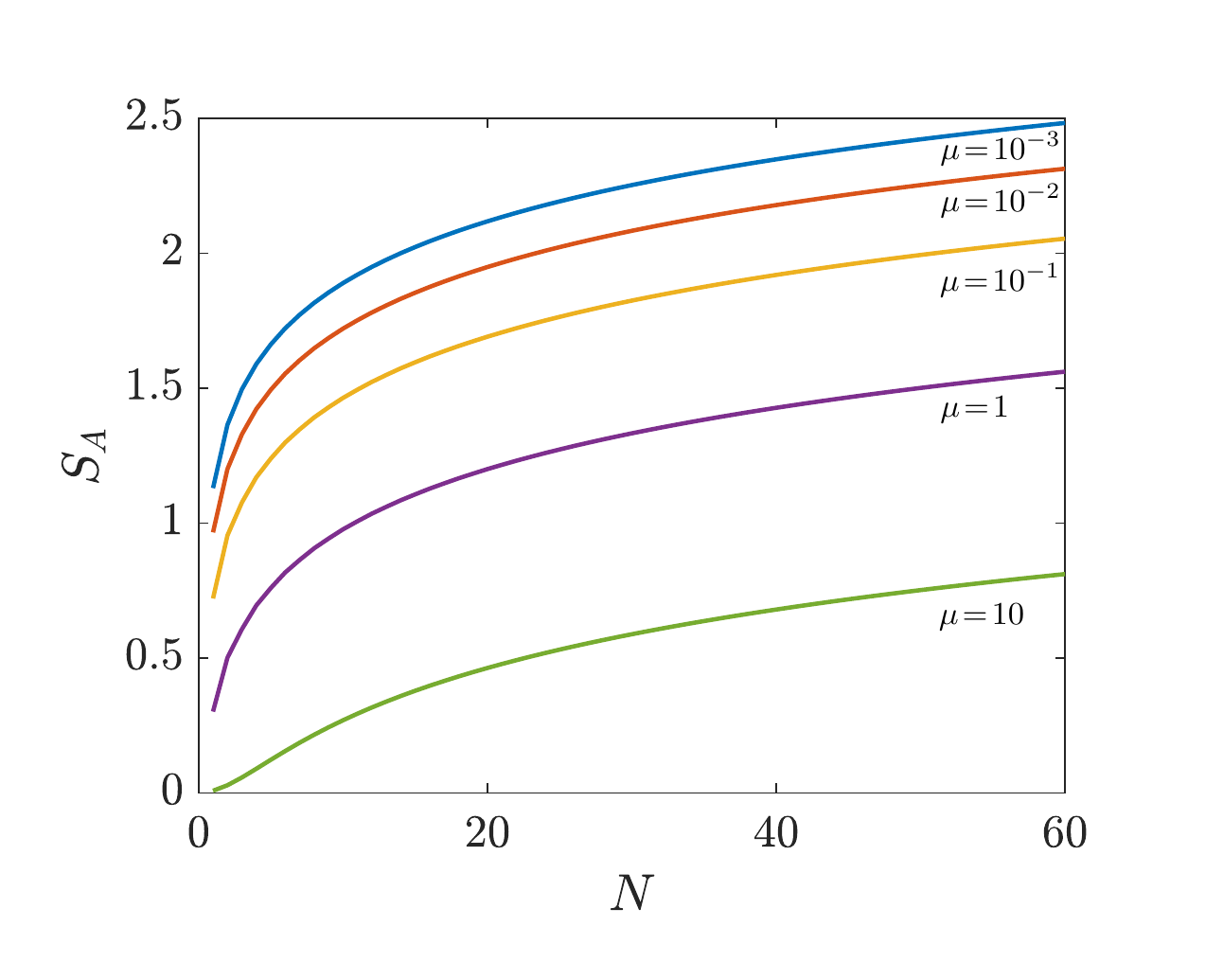}
		\caption{}
		\label{fig:plotEntrp(N)DifMass}
	\end{subfigure}
	\quad
	\begin{subfigure}[t]{0.48\linewidth}
		\includegraphics[width=\linewidth]{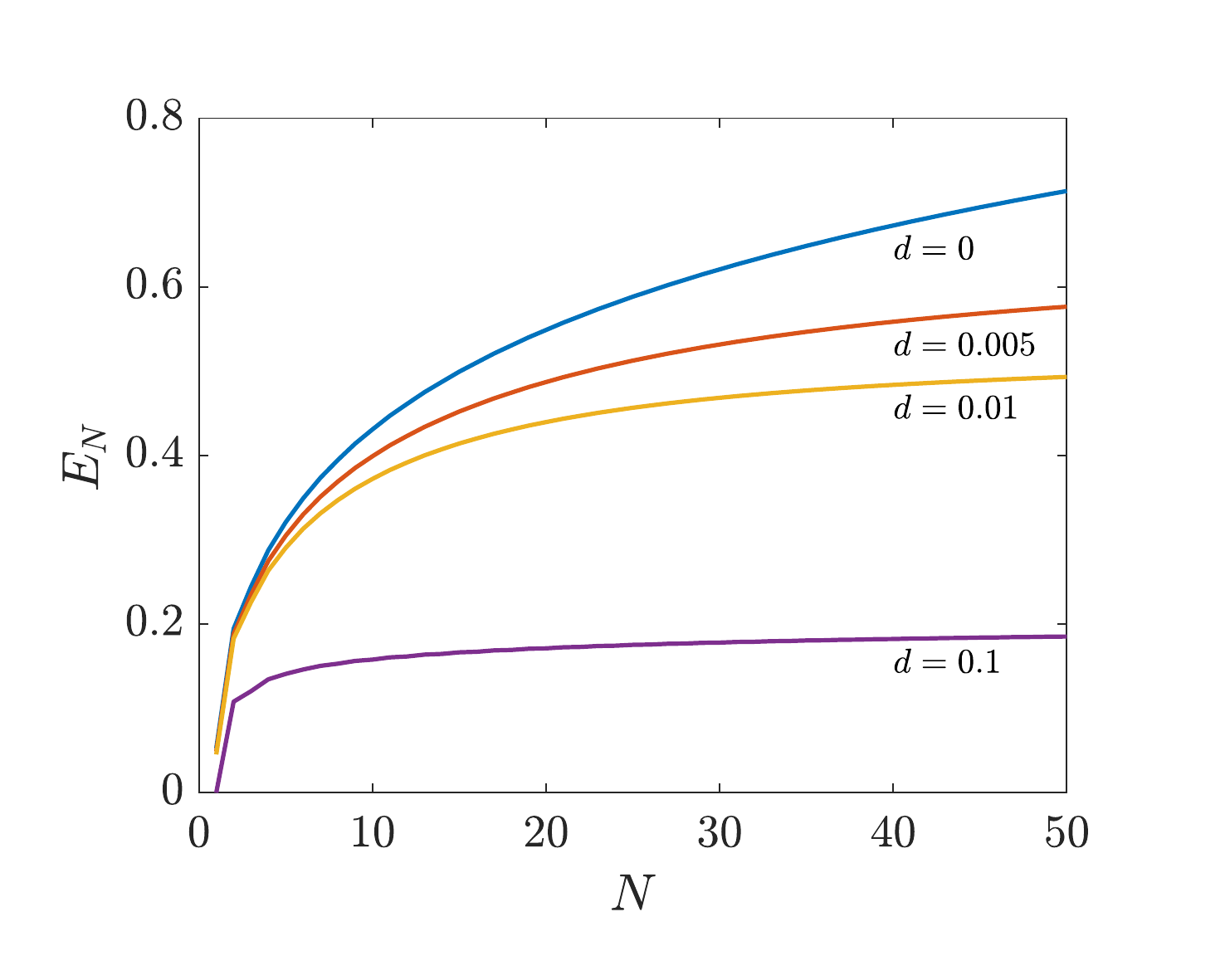}
		\caption{}
		\label{fig:LNeg(N)DifSep}
	\end{subfigure}
	\begin{subfigure}[t]{0.48\linewidth}
		\includegraphics[width=\linewidth]{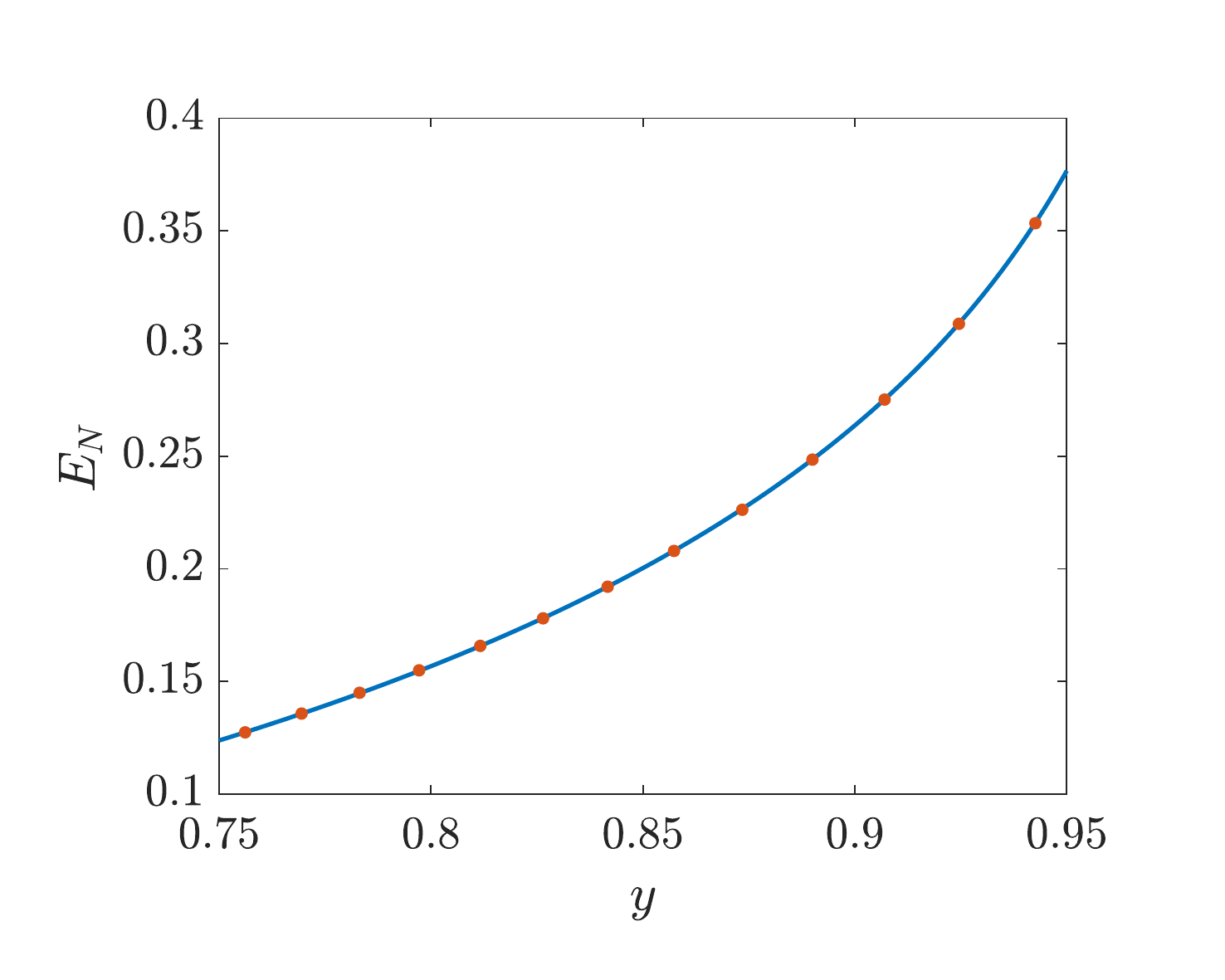}
		\caption{}
		\label{fig:LNeg(y)Flat}
	\end{subfigure}
	\hfill
	\begin{subfigure}[t]{0.48\linewidth}
		\includegraphics[width=\linewidth]{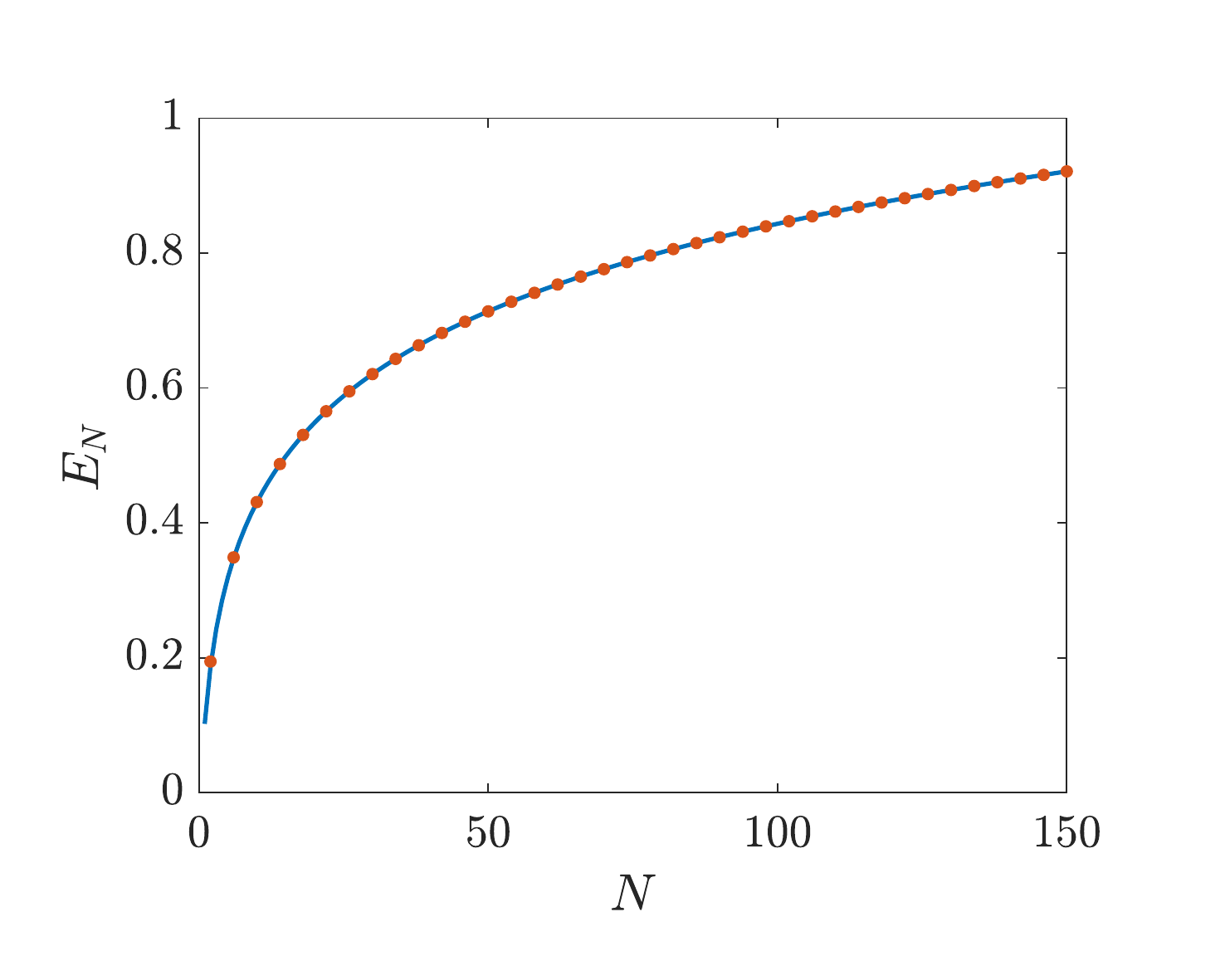}
		\caption{}
		\label{fig:LNeg(N)ZeroSep}
	\end{subfigure}
	\caption{(a) Entanglement Entropy $S_A$ of region A with size $R=1$, as a function of $N$, the number of lowest modes that are included in the covariance matrix. $S_A$  diverges as expected and behaves like $\frac{1}{3}\log{N}+c$ for small masses. (b) The Logarithmic Negativity between the two intervals as a function of the number of lowest modes that are included in the covariance matrix. (c) The Logarithmic Negativity as a function of $y$ (defined in \eqref{eq:41}) for a flat Cauchy surface. The results are fitted to $E_N = -1/4\log(1-y) - 1/2 \log(K(y)) + 0.1615$ (d) The divergence of the Logarithmic negativity as a function of number of lowest modes in the intervals, for a flat Cauchy surface. The intervals are of size 1. The results are fitted to $E_N = -1/4\log(1-(1+a/2 \pi N)^{-2}) - 1/2 \log(K((1+a/2 \pi N)^{-2})) + 0.1615$, $a = 1.176$.}
\end{figure}
In Fig. \ref{fig:plotEntrp(N)DifMass}  we show the entanglement entropy $S_A$ for a single region with $N$ lowest modes in the calculation. It corresponds to the entropy seen by a detector in region A, if it cannot detect particles with energy higher than that of the $N$'th mode. We get that in the massless limit, the entropy fits well with $S_A = \frac{1}{3}\log{N} + c$ which is the same as in the results obtained for a 1+1 scalar field theory~\cite{HOLZHEY1994443, Calabrese_2004, SpatialStructBotero:2004}.

It is interesting to notice that the correct $1/3$ pre-factor appears in our case for any $N_C\geq 1$. This is a somewhat surprising feature, since in the usual spatial-cutoff care, the pre-factor depends on the number of points $n$  within the region $R$, and approaches the value $1/3$ only asymptotically.

\begin{figure}[h!]
	\includegraphics[width=\linewidth, height=0.1\textheight]{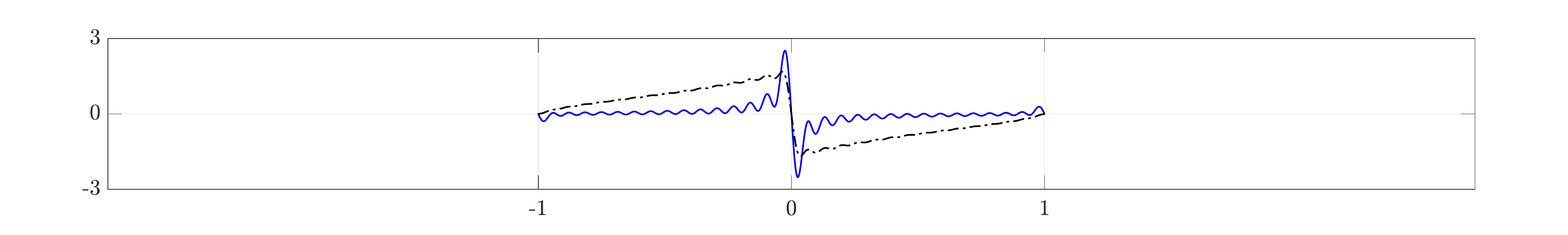}
	\includegraphics[width=\linewidth, height=0.1\textheight]{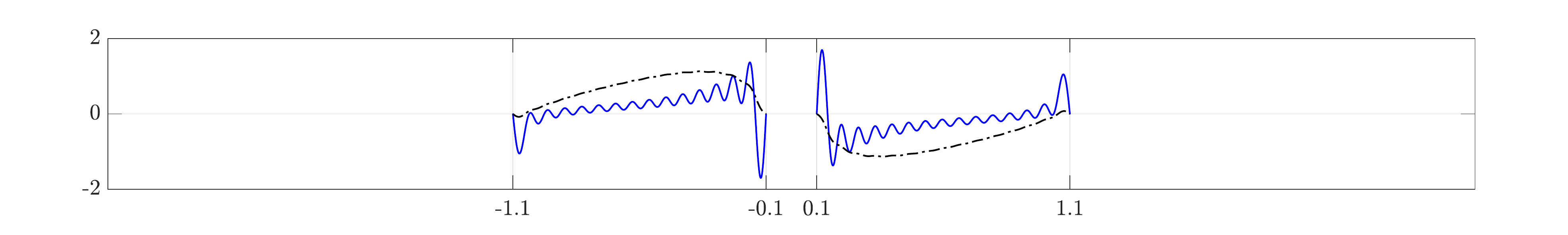}
	\includegraphics[width=\linewidth, height=0.1\textheight]{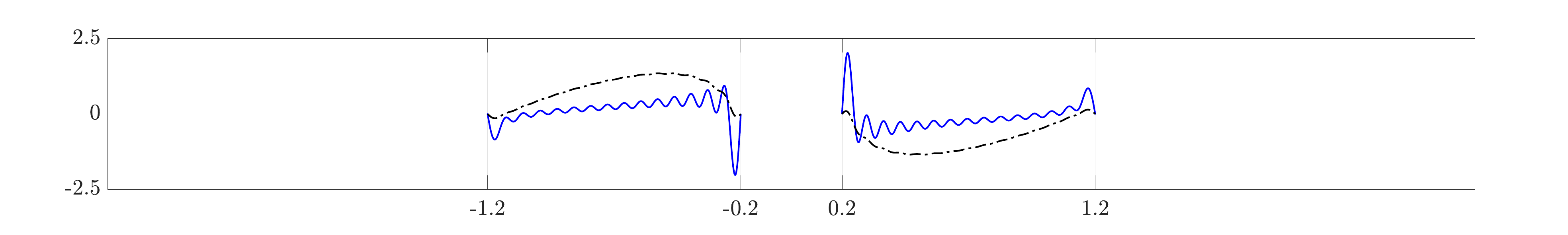}
	\includegraphics[width=\linewidth, height=0.1\textheight]{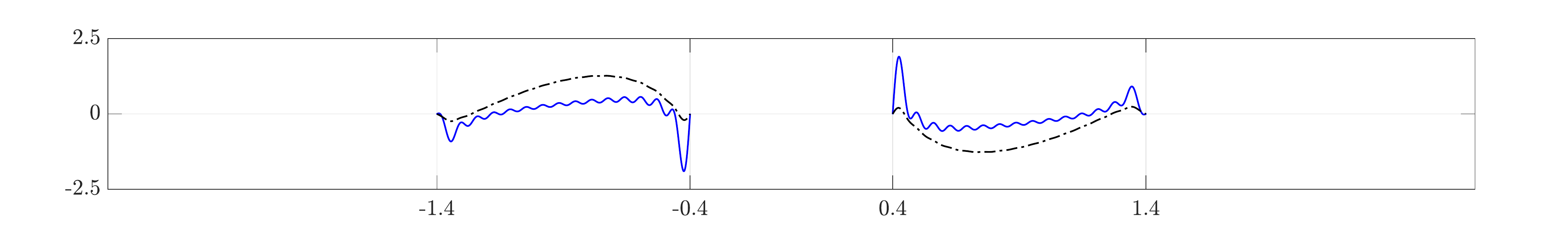}
	\includegraphics[width=\linewidth, height=0.1\textheight]{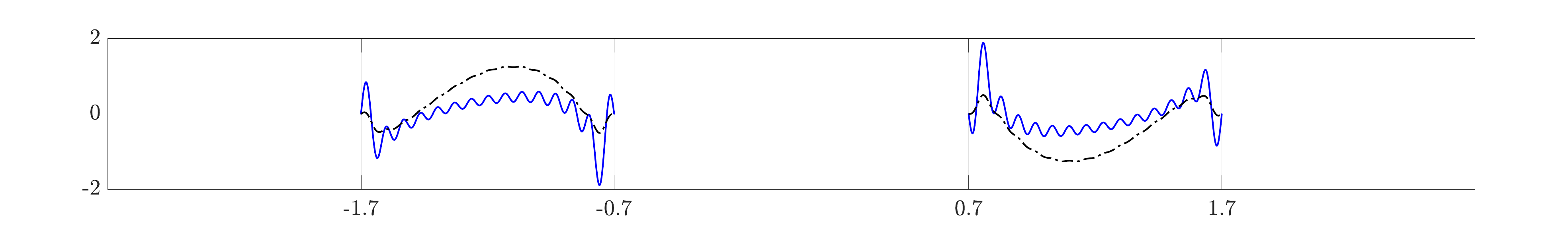}
	\includegraphics[width=\linewidth, height=0.1\textheight]{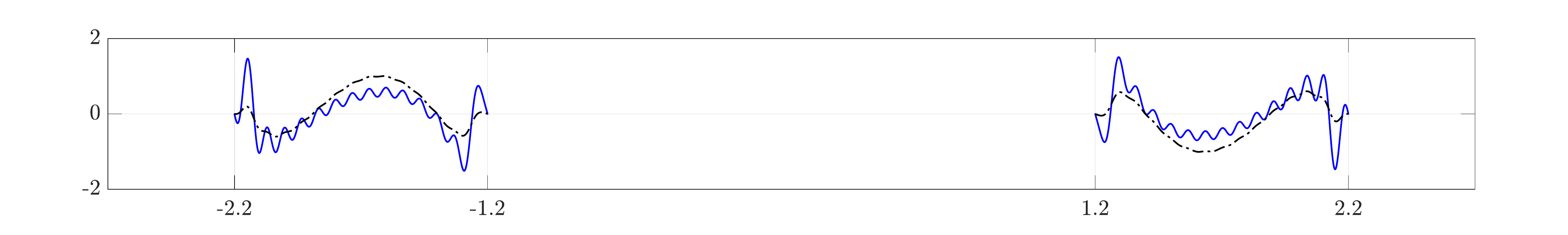}
	\includegraphics[width=\linewidth, height=0.1\textheight]{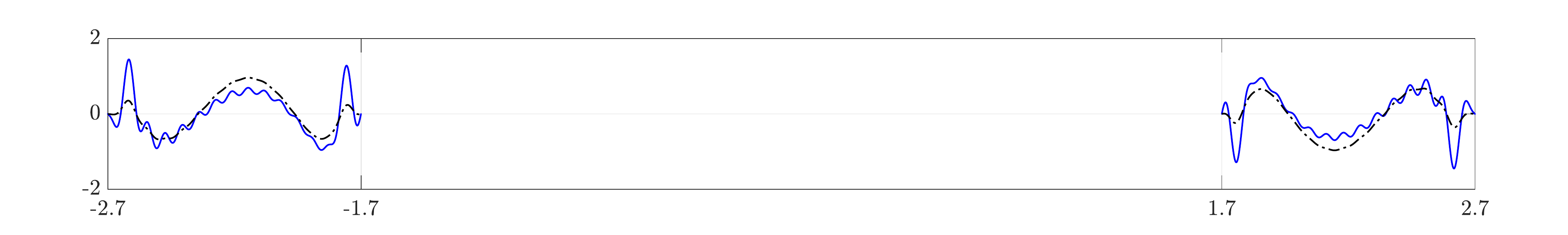}
	\caption[]{The shapes of the most entangled Williamson mode (solid blue for $f_n(x)$, dashed black for $g_n(x)$) for different separations. Each region is taken with $30$ of the lowest modes. The mass equals $10^{-14}/R$.}
	\label{fig: SpatialStructureFlat}
\end{figure}

 In Fig. \ref{fig:LNeg(N)DifSep}, the logarithmic negativity between the two detectors, is plotted for different separations as a function of the number of modes at each detector. We can see the logarithmic negativity diverges when the blocks are not separated. Additionally, we observe that saturation of entanglement is reached when the number of modes at each box is one over the separation, i.e. when the spatial resolution is enough to tell that the regions are separated.
 
 In \cite{EntanglementNegativityinQuantumFieldTheory:2012}, Calbrese et al. obtained that the entanglement is a function of $y$ where
\begin{align}
y = \frac{(x_2-x_1)(x_4-x_3)}{(x_3-x_1)(x_4-x_2)}. \label{eq:41}
\end{align}  
This function diverges as $y$ goes to one, and goes to zero as $y$ vanishes.
For $y$ close to 1, the logarithmic negativity acts as \cite{Calabrese_2013}:
\begin{align}
\epsilon_N = -1/4 \log(1-y) - 1/2\log(K(y))+C. \label{eq:yCloseToOne}
\end{align}
In Fig. \ref{fig:LNeg(y)Flat} we can see an excellent agreement of our results with Eq. \eqref{eq:yCloseToOne}.

As $y$ goes to 1 the entanglement diverge but for a finite number of modes at each interval, i.e. the cutoff, we can say that the effective minimum separation is $\frac{1}{2\pi N}$ instead of 0. If we plug this separation to $y$ we get again an excellent fit with our results, seen in Fig. \ref{fig:LNeg(N)ZeroSep}.

We can also check how does the shape of the modes relate to the entanglement they carry, based on the weight function that are described in Sec. \ref{modesShape}

For the entanglement entropy of a single region, the shapes of the modes are similar to the results in \cite{SpatialStructBotero:2004} except for noisy oscillations that comes from the fact that the number of modes in the region is finite. 
We also note that the mode with the largest symplectic eigenvalue is localized near the edges of the region, and its extremum gets larger as we increase the number of modes in the calculation. This shows that the divergence of the entanglement entropy comes from the boundary of the region.

For the entanglement between two regions, we can see in Fig. \ref{fig: SpatialStructureFlat} how the shape of the most entangled Williamson mode change as the regions are further separated from each other. When the two regions touch, the modes are localized near the edge, but as they separate, the modes need to be shielded from their near environment, and hence we see that they are pushed to the centre.

\section{Milne surfaces} \label{sec:Milne}
\begin{figure}[h!] 
	\centering
	\begin{subfigure}[t]{0.48\linewidth}
		\includegraphics[width=\linewidth]{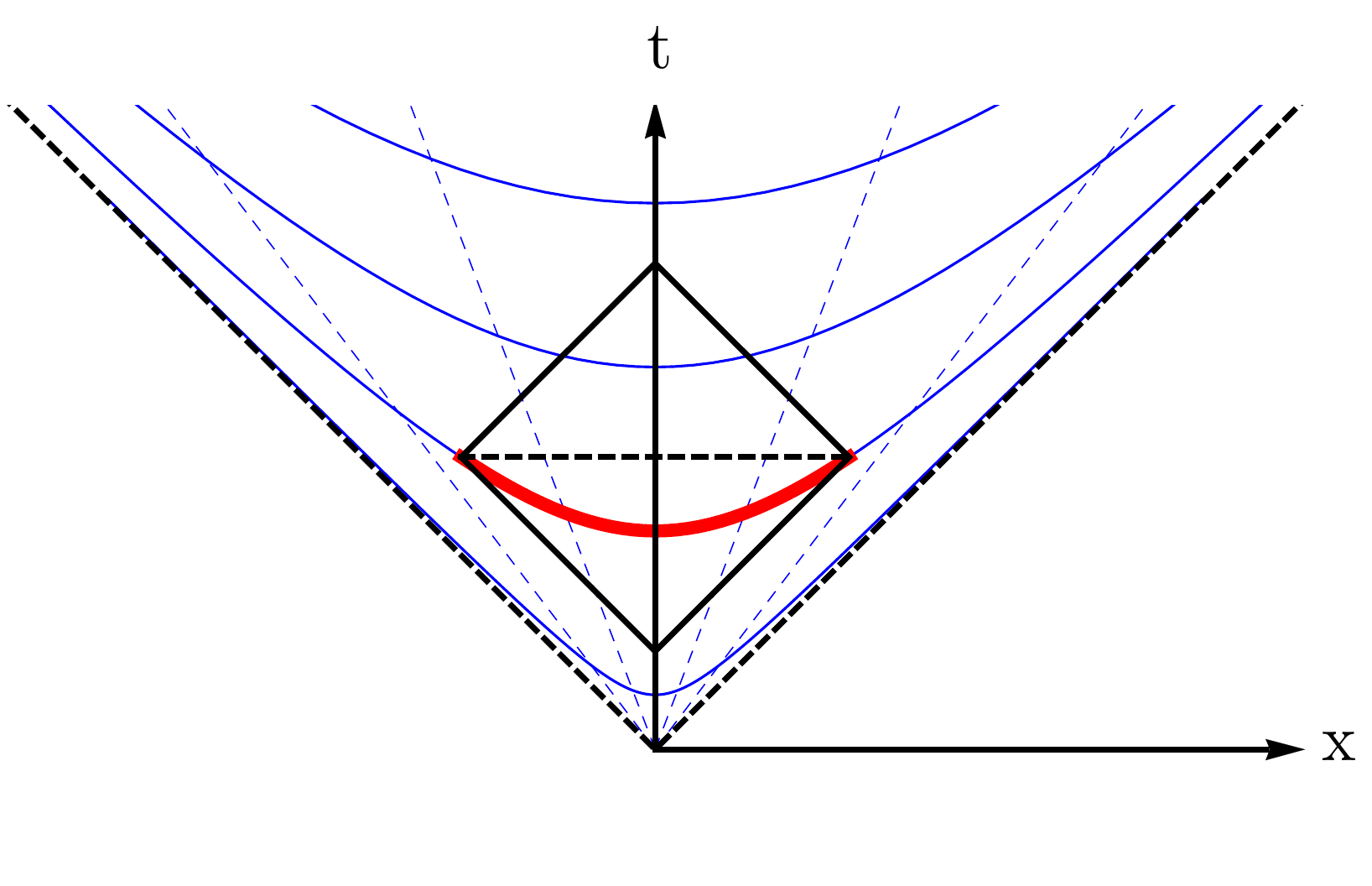}
		\caption{}
		\label{fig:CausalDiamondAndMilneRegion}
	\end{subfigure}
	\hfill
	\begin{subfigure}[t]{0.48\linewidth}
		\includegraphics[width=\linewidth]{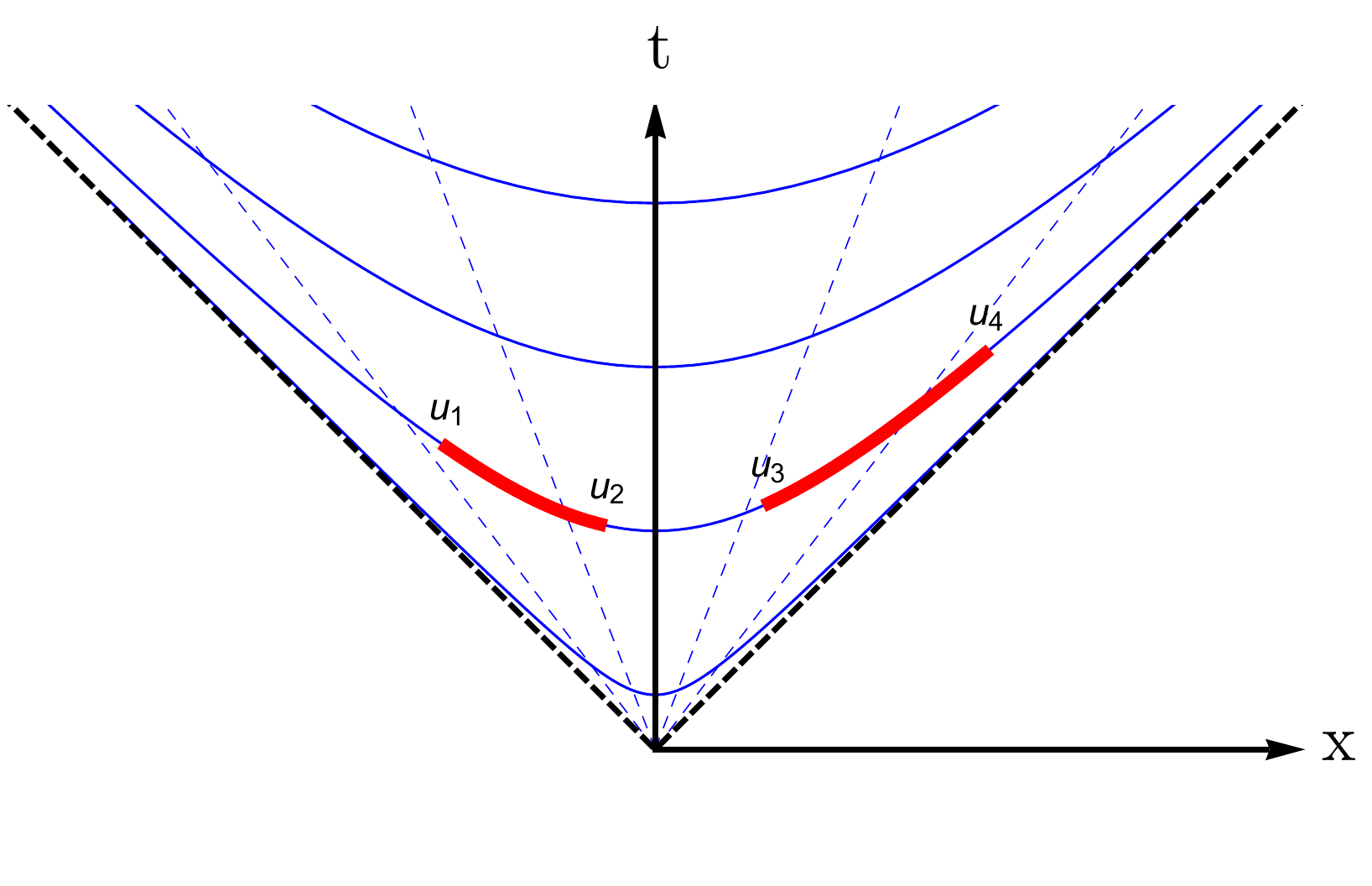}
		\caption{}
		\label{fig:plotintvlsonmilne}
	\end{subfigure}
	\caption{The different configurations we consider for entanglement evaluations on a Milne hypersurface. The blue lines mark hypersurfaces of constant $\eta$, and the dashed lines of constant $z$. (a) The swap is performed at a region embedded on a Milne surface. The entanglement is the same as for a region on a flat surface which shares the same causal diamond. (b) The swap is preformed at two regions on a constant $\eta$ hypersurface. ${u_i}$ are the points in spacetime that mark the regions' ends: $u_i = (t(\eta,z_i),x(\eta,z_i))$} 
	\label{fig:bothMilneConfig}
\end{figure}

The Milne coordinates are defined by $x = {\mathrm{e}^{a \eta} \over a}\sinh(az), t = {\mathrm{e}^{a \eta} \over a} \cosh(a z) $, with the metric $ds^2 = \mathrm{e}^{2a \eta}(dz^2-d\eta^2)$. Surfaces of constant $\eta$ and $z$ can be seen in Fig. \ref{fig:bothMilneConfig}.
In this section we evaluate the Minkowski vacuum's entanglement between regions on a hypersurface $\sigma$, defined by a constant $\eta$. Interval A will be between $(z_1, z_2)$ and interval B between $(z_3, z_4)$. A single and two intervals, are illustrated in Fig. \ref{fig:CausalDiamondAndMilneRegion} and Fig. \ref{fig:plotintvlsonmilne}.

The momentum operator along $\sigma$ is $\pi(x) = {\partial \phi(x) \over \partial \eta}$. The discretizer operators after the swap become:

\begin{align}
\psi^B_n = \sqrt{{2}\over{\Delta_B}} \int^{z_4}_{z_3}dz\phi(x(\eta,z))\sin(\frac{\pi n}{ \Delta_B}(z-z_3)),  \label{eq:43} \\
\pi^B_n = \sqrt{{2}\over{\Delta_B}} \int^{z_4}_{z_3}dz\partial_\eta \phi(x(\eta,z))\sin(\frac{\pi n}{ \Delta_B}(z-z_3)), 
\end{align}
where $\Delta_B = z_4-z_3$.
Let us take $a$ to be equal to $1$. For small masses, the CM elements will be:
\begin{align}
& M_{\psi^A_n,\psi^B_m} =  {-2\over \pi \sqrt{\Delta_A \Delta_B}}\int_{0}^{\Delta_A}dz\int_{0}^{\Delta_B}dz'\sin\left({\pi n\over \Delta_A}z\right)\sin\left({\pi m\over \Delta_B}z'\right)\log\left(\left|2 M_0 \mathrm{e}^{\eta}\sinh\left(\frac{z-z'+z1-z3}{2}\right)\right| \right), \\
& M_{\pi^A_n\pi^B_m} = {-2\over \pi \sqrt{\Delta_A \Delta_B}}\int_{0}^{\Delta_A}dz\int_{0}^{\Delta_B}dz'\sin\left({\pi n\over \Delta_A}z\right)\sin\left({\pi m\over \Delta_B}z'\right)\frac{1}{4\sinh^2\left(\frac{z-z'+z1-z3}{2}\right)}, \label{MpiMilne}
\end{align}   
where $\Delta_A = z_2-z_1$. 
We can see that these equations are Lorentz invariant as they depend on differences in the z axis. Additionally, for vanishing masses, they do not depend on $\eta$.
We can integrate by parts Eq. \eqref{MpiMilne} to get:
\begin{align}
& M_{\pi^A_n\pi^B_m} = {-2\pi nm\over \Delta_A \Delta_B \sqrt{\Delta_A \Delta_B}}\int_{0}^{\Delta_A}dz\int_{0}^{\Delta_B}dz'\cos\left({\pi n\over \Delta_A}z\right)\cos\left({\pi m\over \Delta_B}z'\right)\log\left(\left|\sinh\left(\frac{z-z'+z1-z3}{2}\right)\right| \right),
\end{align} 
and then, by finding a vector $-u(z,z')dz+u(z,z')dz'$ who's curl equal the integrand, we can use green's theorem to simplify the numeric integration.  

To check how does the entanglement change with the region's size, we increase the region with $N$, the number of lowest modes in the covariance matrix such that $\Delta_A = \Delta N$, where $\Delta$ is constant. This insures that the wavelength of the highest mode stays constant and what that increases with the region's size, is the number of times this wavelength enters in the region. This corresponds in lattice discretization to holding the lattice space constant and increasing the number of lattice points of the region. We find that for $\Delta < 1$ the entanglement entropy of a single region centred around the origin is:
\begin{align}
S_A = \frac{1}{3}\log(\frac{2\tau \sinh(\frac{\Delta}{2} N)}{\tau\kappa})+c = \frac{1}{3}\log(\frac{R}{\epsilon}) +c, \label{eq:entrpyCurved}
\end{align}
where $\kappa$ is some constant that serves as a dimensionless ultraviolet cutoff, $\epsilon = \tau \kappa$ controls the spatial resolution, and $\tau = \text{e}^\eta$. This result was also obtained in \cite{Berges:2018aa} and is similar to the result on the flat surface. This suggests that the entanglement does not depend on deformations inside the casual diamond. For $\Delta>1$ the factor of the logarithm decreases as $\Delta$ gets larger. 

We can check how the entanglement changes with the effective cutoff when holding $\Delta_A$ constant while increasing $N$. This corresponds to increasing the spatial resolution while keeping the region's size constant. We get that for $\Delta_A<1$, $\epsilon = 2\tau\sinh(\frac{\Delta_A}{2 N})$ as expected, as for small $\Delta_A$ the region is in a part of the surface that is approximately flat. For $\Delta_A>1$ the fit is not as good due to data points with small $N$. 

This shows that for spatial resolutions higher then $\tau$, Eq. \eqref{eq:entrpyCurved} holds and the entanglement is indeed the same in any region inside the casual diamond. For $\epsilon$ larger then $\tau$ we suspect that the deviation from Eq. \eqref{eq:entrpyCurved} is because the sin modes, as defined in \eqref{eq:43}, are not modes that diagonalize the Hamiltonian on that surface thus the effective energy cutoff is not proportional to N. In addition, it is only reasonable that we will not notice the true entanglement when the effective cutoff is less than $\tau$, as the degrees of freedom that are not traced, are not continuous with respect to the curvature.

For the case of two separated regions, if we extend the definition of $y$, the four point function to
\begin{align}
y\rightarrow \frac{||u_2-u_1|| \  ||u_4-u_3||}{||u_4-u_2|| \ ||u_3-u_1||}
\end{align} 
where $u_i$ is a point in space time and $||*||$ is the Minkowski spacetime interval, we get that the logarithmic negativity is the same function of $y$ for a flat surface, found in \cite{Calabrese_2013}. Equation \eqref{eq:41} is just the special case of $y$ on a flat region. We can express $y$ in terms of the flat and Milne coordinates on the Milne surface:
\begin{align}
y = & \frac{\sqrt{(\Delta_{21})^2 - (\sqrt{\tau^2+x_2^2}-\sqrt{\tau^2+x_1^2})^2}\sqrt{(\Delta_{43})^2 - (\sqrt{\tau^2+x_4^2}-\sqrt{\tau^2+x_3^2})^2}}{\sqrt{(\Delta_{42})^2 - (\sqrt{\tau^2+x_4^2}-\sqrt{\tau^2+x_2^2})^2}\sqrt{(\Delta_{31})^2 - (\sqrt{\tau^2+x_3^2}-\sqrt{\tau^2+x_1^2})^2}} \label{eq:50} \\ \nonumber = & \frac{\sinh(\frac{z2-z1}{2})\sinh(\frac{z4-z3}{2})}{\sinh(\frac{z4-z2}{2})\sinh(\frac{z3-z1}{2})} 
\end{align}

\section{Remarks on general hypersurfaces and lightcone entanglement} \label{sec:Remarks}
The results in the previous section are in agreement with the notion that the entanglement does not depend on deformations inside the casual diamonds of the regions, and on the shape of the hypersurface.
In the definition above, $y$ is the quotient of proper distances along straight lines in Minkowski spacetime. There is no reminiscence to the fact that the entanglement was evaluated on Milne hypersurfaces. We can conclude that this is a feature of the Minkowski vacuum and true for any surface we choose. 
 
In the definition above, $y$ is the quotient of proper distances along straight lines in Minkowski spacetime. There is no reminiscence to the fact that the entanglement was evaluated on Milne hypersurfaces 

 \begin{figure}[h!]
 	\centering
 	\includegraphics[width=0.5\linewidth]{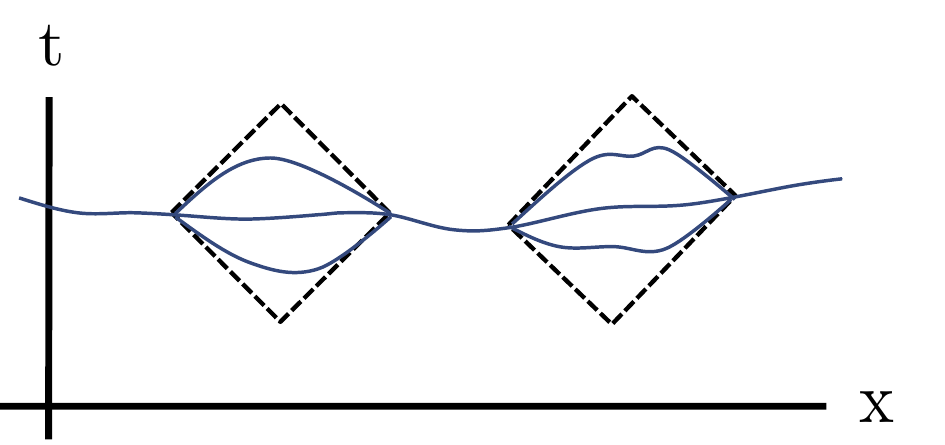}
 	\caption[]{Different hypersurfaces that intersect at the endpoints of the causal diamonds of certain regions. The entanglement between regions does not depend on the shape of the hypersurface, but only on the endpoints of the regions }
 	\label{fig:plot2RegionsDeform}
 \end{figure}

\begin{figure}[h!]
	\centering
	\includegraphics[width=0.9\linewidth]{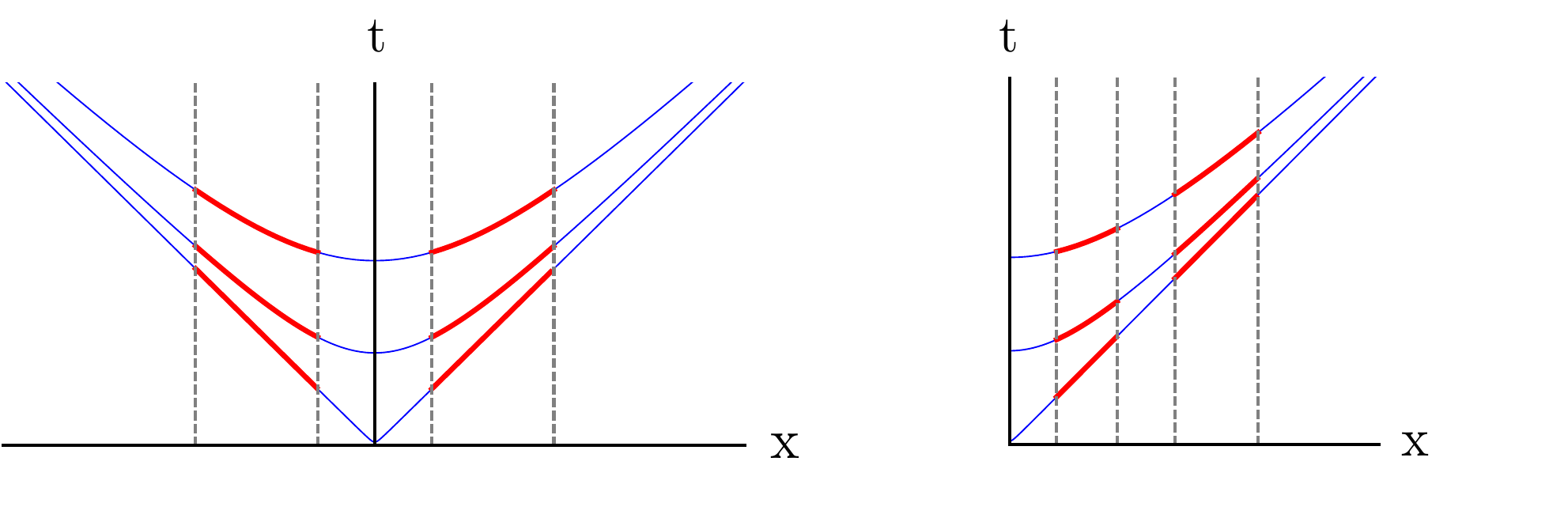}
	\caption[]{We consider sections with a constant separation on the $x$ axis that approach the light cone. (Left) the sections are symmetric on either sides of the light cone, getting disentangled as they approach it. (Right) the sections are on the same side of the light cone and stay entangled.     }
	\label{fig:plotLightCone}
\end{figure}

Let us consider some interesting scenarios. If we take two infinite regions that are separated on the Milne surface, meaning the limit of $z_1 = -\infty, z_4 = \infty$ in Eq. \eqref{eq:50}, we get that $y$ does not approach one, hence the entanglement between the regions is finite. This is unlike two separated infinite regions on a flat surface for which the entanglement diverges. We can also hold the separation between them in the $x$ axis constant and decrease $\tau$ so that the two regions approach the light cone (as depicted on the left plot of Fig. \ref{fig:plotLightCone}). We get that $y$ goes to zero and the entanglement vanishes, showing that the two sides of the light cone are not entangled if separated by a finite region. 

Further more, we can look at two finite regions on the same side of the light cone. We take two regions with $x_1>0$ and decrease $\tau$ while holding the positions on the $x$ axis (depicted on the right plot of Fig. \ref{fig:plotLightCone}). In this limit, $y$ is the same as in Eq. \eqref{eq:41}: the spacetime intervals are replaced with the lengths of the projections of those regions on the $x$ axis. To derive this result it is enough to use the result on a flat surface by \cite{EntanglementNegativityinQuantumFieldTheory:2012} for which $y$ is defined as in Eq. \eqref{eq:41} and assume the Lorentz invariance of the vacuum. 

On the other hand, from the vacuum's Lorentz invariance and the results on a flat surface for which the entanglement is proportional to $\log(R)$, we might expect that the entanglement entropy of a region on the lightcone is zero. How can it be then, that a region on the lightcone is not entanglement with its environment but has non zero negativity with a distant region? To claim that the entanglement entropy depends on the region's size, we need to add an energy cutoff to the theory, but with it, the vacuum is no longer Lorentz invariant. In order for the vacuum to be invariant we need to look at infinite degrees of freedom that will give a diverging entanglement entropy for any region in spacetime.

It seems that although deforming the surface inside the regions does not effect the total entanglement, it might change the arrangement of the degrees of freedom that are entangled. Consider again the two regions with constant separation, symmetric around the $t$ axis (Left plot on Fig. 7). As the regions approach the light cone, the entanglement between them vanishes. What happens to the entangled degrees of freedom in this process? 
If we look at the direction of the vector that is orthogonal to the Milne surface, we see that it points to the origin. This is the direction of time on that surface. From that we can speculate that the density of the degrees of freedom that are responsible for the entanglement on the flat regions are now concentrated in the middle of the lightcone wedge. 

If this is the case, we can further speculate on the entangled degrees of freedom of two light cone wedges that are next to each other (such as the hypersurfaces that approach the dashed lines in the causal diamonds of Fig. \ref{fig:plot2RegionsDeform}). While the degrees of freedom that are entangled to the environment are at the endpoints, those that are entangled between the regions are now localized at the centre. This corresponds to the Williamson modes' shapes we saw in Sec. 7, which are not localized at the edges. When deforming the flat regions to the light cone wedges, the modes are squeezed to the centre.

\section{Comparison to lattice discretization} \label{sec:compLattice}

Theories on lattices are approximations to continuous ones such that there are no degrees of freedom between the lattice spacing. Consequently, the entropy of a subregion on a lattice measures the entanglement between all the degrees of freedom inside to all the ones outside. This differs from the present approach in such that the cutoff comes not from removing degrees of freedom from the theory but from tracing them over. When looking at the entanglement of $N$ modes of the discretizer, not only that the degrees of freedom of the system's field are traced but also modes of the discretizer that are larger than $N$. Hence the entanglement is also between these modes and not only between the complement of the region. In sum, while the entanglement entropy between complement regions on a lattice is the entropy of a subregion of a discretized theory, in the discretizer method, it is the entropy a detector can see if it restricted to this subregion and cannot measure modes higher than $N$, in the continuous theory. For the case of two separate regions, the entanglement does not diverge with the cutoff. Hence, the finite entanglement, between the two regions will be the same in the two approaches.

Although the difference in these approaches, we see from numeric calculations on flat hypersurfaces, that the entanglement between the first $N$ modes of the discretizer to the rest of the discretizer modes, saturates when $N$ grows. We also observe that the entanglement between $N_A$ lowest modes of region $A$ to $N_B$ lowest modes of a region next to $A$ saturates as a function of $N_B$, around $N_B = N_A$. This suggest that most of the entanglement comes from the contribution of modes smaller than $N_A$ in the theory, which helps explain why the results for the entanglement entropy in Sec. \ref{sec:flat}, agree with those of lattice calculations and the analytical results from conformal field theory. 

In fact, it appears from Sec.~\ref{sec:flat} that the discretizer approach has an advantage over a lattice theory in converging faster to the analytical results of the continuous case. This, we assume, is because the starting ground of the discretizer is the continuous theory while the degrees of freedom in a region of a lattice appears continuous only for a small enough cutoff. 

On a different note, the physical meaning addressed to the effective cutoff that comes from tracing over all but $N_c$ modes, depends on the choice of orthonormal functions $h_n(z)$ of the discretizer expansion. For flat metrics in any dimension, one can choose sine functions and set $N_c$ such that the minimum wavelength in any direction is $a$. This naturally corresponds to a lattice with the same spacing. In $3$ spatial dimensions for example, this should reproduce $S_A \sim A/a^2$, where $A$ is the area of the region. Different choices of $h_n(z)$, and direction dependent cutoffs will correspond to a lattice with different arrangements and spacings. However, a possible way to relate $N_c$ to a physical distance is to compute the entanglement between regions separated by a distance  $r$. The entanglement will saturate for $N_c(r)$. Another possibility is to relate the entropy to the change in energy of the discretizer. The discretizer's Hamiltonian can be expressed in the terms of $\psi_n$ and $\pi_n$, which might not be diagonal. The difference between the Hamiltonian of $N_c$ modes' expectation value, before and after the swap, will give the energy needed to produce such entanglement.

\section{Conclusions} \label{sec:conclusions}
We presented a covariant scheme, that is appropriate for studying entanglement between arbitrary regions on 
general hypersurfaces. 
To this end, we replaced the UDW point-like detector, with 
a relativistic "discretizer field", and introduced a covariant interaction that fully swaps the system's and discretizer's states, within the regions. The method
provides a natural cutoff bypassing difficulties that arise when imposing a spatial discretization. 

We applied the approach in several toy examples and computed  the entanglement of complementary and separated regions in 1+1 dimensions, flat and Milne hypersurfaces, in the Minkowski vacuum.
In the flat case, our results corroborate with previous works.
For the Milne hypersurface we obtained, that entanglement remains invariant under local deformations inside the causal diamond, and depends only on the boundary.

The method also provides  a way to examine the spatial structure of entanglement within particular regions.  This "internal" entanglement structure 
is manifested for each region by the 
shape of the Williamson modes, with respect to their relative partial contribution to entanglement. 
We find that as regions become more separated, entanglement arises from modes that are localized further away from the boundaries.  
It is interesting to note, that while that total entanglement 
remains unchanged by local deformations, the "internal" entanglement structure depends on the shape of the hypersurface.

\section{Appendix: Relativistic von-Neumann measurements}

We begin by formulating von-Neumann measurements, 
in the framework of a relativistic quantum field theory. 
In the simplest case, the system and measuring device, will be represented by two relativistic scalar fields, $\phi(x)$ and $\Phi(x)$, respectively, in a Minkowskian spacetime. 

In the Hamiltonian formalism, each field has a conjugate momenta $\pi$ and $\Pi$,  that satisfy the ordinary canonical commutation relations,  $[\phi(x),\pi(x')]= i \delta(x-x')$, on hypersurfaces with $t=const$.
The systems are initially uncoupled and described by the free relativistic Hamiltonians, $H_0=H_S+H_D$.

Next we add a measurement interaction
that is temporarily ``switched on'' between 
the detector and system, and designed to 
couple with the relevant to-be-measured field observable. To measure $\phi(x)$, consider the interaction Hamiltonian
\begin{align}
H_I = -\int d^n x  g(x,t) \Phi(x,t) \phi(x,t).
\end{align} 
It is non-zero, only in a spacetime regions wherein 
the coupling function $g(x,t)\neq 0$.
In the limit of an instantaneous measurement, $g(x,t)= f(x)\delta(t)$. 

Integrating Hamilton's eqs.  we have
\begin{align}
\delta \Pi(x) = f(x)\phi(x,0) \label{eq:appendMeas}
\end{align}
where $\delta \Pi(x)=\Pi(x,t) - \Pi(x,0^-) $, is the local change of the field pointer in the limit $t\to 0^+$.
The proportionality of the shift to the observable, 
motivates the familiar terminology, referring 
to  $\Pi$ as the "pointer" (in fact, one pointer, per each spacetime point in the present case).

It will be helpful to consider 
an averaged field pointer 
$\Pi_D= \int \Pi(x)d^nx$  
\begin{align}
\delta \Pi_D = \int d^nx f(x)\phi(x)
\end{align}
Rather then measuring the field at a point, something that often diverges, $\Pi_D$  couples to a field distribution,  that is, the field smeared with (our choice) of the "test function" $f(x)$.  
Smeared fields are better behaved mathematical objects, and often used 
within Algebraic field theory~\cite{HaagsBook}.

It is well known that, unlike the case of non-relativistic quantum theories, special relativity limits the set of observables that can be measured instantaneously without conflicting with causality \cite{Aharonov81, Aharonov86, Sorkin:1993gg, Preskill, GoismanReznikSemilocal}.
In the following, we must therefore
keep our formalism covariant. 

It is easy to check that the measurement interaction Hamiltonian is local and Lorentz invariant, by recasting it in its corresponding Lagrangian,
\begin{align}
{\cal L} = g(x)\Phi(x)\phi(x).
\end{align}

For a complete measurement scheme, we need to include ``velocity`` measurements. Our field has a "velocity" operator, that is related to the conjugate field $\pi=\partial_t\phi$. This then suggests a velocity measurement of a form 
\begin{align}
H_I =-\int d^n x g(x,t) \Phi(x,t) \pi(x,t)
\end{align}
that seems at first sight, as a harmless extension of the former field case.  
Notice however, that while $\phi$ and $\Phi$ transform like scalars, their conjugates $\pi$ and $\Pi$, do not. 

The Lagrangian in this case is
\begin{align}
{\cal L} = \int d^nx \biggl( g(x)\Phi(x)\partial_t \phi(x) + {1\over2}g(x)^2 \Phi(x)^2  \biggr)
\end{align}
The first term, on the right-hand-side, can be written as $ g(x)\Phi(x) \epsilon^\mu \partial_\mu \phi(x)$, where $\epsilon^\mu$ is a four-vector orthogonal to the hypersurface $t=0$, and is hence invariant. However, unlike the former case, there appears an extra negative spring-like term $\sim g^2 \phi^2$, as discussed in Sec. \ref{sec:swap}.

\bibliography{CurvedEntanglementBib}{}
\bibliographystyle{ieeetr}
\end{document}